\documentclass[prb,aps,superscriptaddress,showpacs,twocolumn]{revtex4-1}

\usepackage{graphicx}
\usepackage{epstopdf}

\begin{document}

\title{Evidence for anisotropic polar nanoregions in relaxor Pb(Mg$_{1/3}$Nb$_{2/3}$)O$_3$: A neutron study of the elastic constants and anomalous TA phonon damping}

\author{C. Stock}
\affiliation{NIST Center for Neutron Research, National Institute of Standards  and Technology, Gaithersburg, Maryland 20899-6100, USA}
\affiliation{Indiana University, 2401 Milo B. Sampson Lane, Bloomington, Indiana 47404, USA}
\author{P. M. Gehring}
\affiliation{NIST Center for Neutron Research, National Institute of Standards  and Technology, Gaithersburg, Maryland 20899-6100, USA}
\author{H. Hiraka}
\affiliation{Institute for Materials Research, Tohoku University, Sendai, 980-8577, Japan}
\author{I. Swainson}
\affiliation{National Research Council, Chalk River, Ontario, Canada K0J 1J0}
\author{Guangyong Xu}
\affiliation{Condensed Matter Physics and Materials Science Department, Brookhaven National Laboratory, Upton, New York 11973-5000, USA}
\author{Z.-G. Ye}
\affiliation{Department of Chemistry and 4D Labs, Simon Fraser University, Burnaby, British Columbia, Canada V5A 1S6}
\author{H. Luo}
\affiliation{Shanghai Institute of Ceramics, Chinese Academy of Sciences, Shanghai 201800, China}
\author{J. -F. Li}
\affiliation{Department of Materials Science, Virginia Tech., Blacksburg, Virginia 24061, USA}
\author{D. Viehland}
\affiliation{Department of Materials Science, Virginia Tech., Blacksburg, Virginia 24061, USA}

\date{\today}

\begin{abstract}

We use neutron inelastic scattering to characterize the acoustic phonons in the relaxor Pb(Mg$_{1/3}$Nb$_{2/3}$)O$_3$ (PMN) and demonstrate the presence of a highly anisotropic damping mechanism that is directly related to short-range, polar correlations.  For a large range of temperatures above T$_c \sim 210$\, K, where dynamic, short-range, polar correlations are present, acoustic phonons propagating along [1$\overline{1}$0] and polarized along [110] (TA$_2$ phonons) are overdamped and softened across most of the Brillouin zone.  By contrast, acoustic phonons propagating along [100] and polarized along [001] (TA$_1$ phonons) are overdamped and softened for only a limited range of wavevectors, $q$.  The anisotropy and temperature dependence of the acoustic phonon energy linewidth $\Gamma$ are directly correlated with the neutron elastic diffuse scattering cross section, indicating that polar nanoregions are the cause of the anomalous behavior.  The damping and softening vanish for $q \rightarrow 0$, i.\ e.\ for long-wavelength acoustic phonons near the zone center, which  supports the notion that the anomalous damping is a result of the coupling between the relaxational component of the diffuse scattering and the harmonic TA phonons.  Therefore, these effects are not due to large changes in the elastic constants with temperature because the elastic constants correspond to the long-wavelength limit.  We compare the elastic constants we measure to those from Brillouin scattering experiments and to values reported for pure PbTiO$_3$.  We show that while the values of C$_{44}$ are quite similar, those for C$_{11}$ and C$_{12}$ are significantly less in PMN and result in a softening of (C$_{11}$-C$_{12}$) over PbTiO$_3$.   The elastic constants also show an increased elastic anisotropy (2C$_{44}$/(C$_{11}$-C$_{12}$)) in PMN versus that in PbTiO$_3$.  These results are suggestive of an instability to TA$_2$ acoustic fluctuations in PMN and other relaxor ferroelectrics.  We discuss our results in the context of the current debate over the ``waterfall" effect and show that they are inconsistent with acoustic-optic phonon coupling or other models that invoke the presence of a second, low-energy, optic mode.

\end{abstract}

\pacs{74.72.-h, 75.40. Gb}

\maketitle

\section{Introduction}

	Relaxor ferroelectrics have received a great deal of attention from the physics and materials science communities because of their exceptional dielectric properties and potential for use as piezoelectric devices.~\cite{Park97:82,Cowley11:60,Xu10:79}  Pb(Mg$_{1/3}$Nb$_{2/3}$)O$_3$ (PMN), Pb(Zn$_{1/3}$Nb$_{2/3}$)O$_3$ (PZN), and Pb(Zr$_{1-x}$Ti$_{x}$)O$_{3}$ (PZT) are prototypical relaxors that display a broad and frequency-dependent peak in the dielectric response.~\cite{Ye98:81,Ye09:34}   Single crystals of solid solutions between PMN, PZN, or PZT and ferroelectric PbTiO$_3$ (PT) exhibit piezoelectric coefficients and electro-mechanical coupling factors that are much larger than those of conventional piezoelectric materials.  Despite many theoretical and experimental studies of these and other relaxor ferroelectrics, neither the relaxor ground state nor the relaxor transition are well understood, and there are few materials with comparable dielectric properties.~\cite{Hirota06:75}

\begin{figure}[t]
\includegraphics[width=80mm]{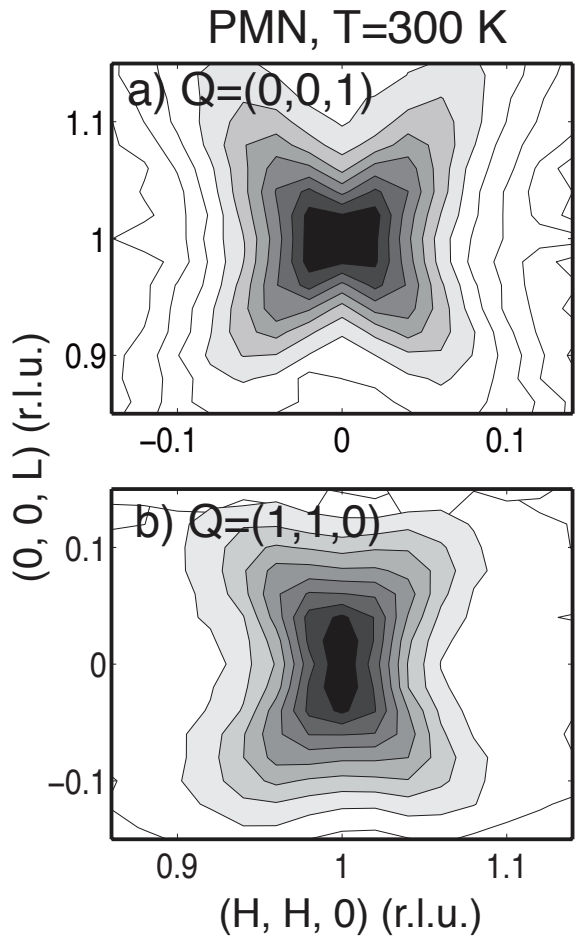}
\caption{Elastic diffuse scattering intensity contours measured near $a)$ $\vec{Q}=(0,0,1)$ and $b)$ $\vec{Q}=(1,1,0)$ at $T=300$\,K on SPINS ($E_f=4.5$\,meV).  The sample (S-II) was aligned in the (HHL) scattering plane. }
\label{diffuse_summary}
\end{figure}

	The structural properties of PMN and PZN display several common features.  At a temperature $T_d$, far above the critical temperature $T_C$ below which long range ferroelectric order can be induced in both PMN and PZN by cooling in a sufficiently strong external electric field, evidence of the formation of local regions of polar order, also known as polar nanoregions, has been obtained from refractive index measurements and neutron scattering pair distribution function analysis.~\cite{Burns83:48,Jeong05:94}  The existence of such locally ordered regions implies that short-range, polar correlations are present, and these are manifested in the form of strong neutron and x-ray diffuse scattering intensity located in the vicinity of Bragg peaks (Ref.~\onlinecite{Vak95:37,Hirota02:65,Welberry05:38}). The geometry of this diffuse scattering in reciprocal space is illustrated for two different Brillouin zones in PMN at room temperature in Fig.~\ref{diffuse_summary} (identical results have been obtained for PZN as well).  In contrast to the sharp, resolution-limited, Bragg peaks that characterize long-range ordered structures, the diffuse scattering originating from short-range, polar correlations is broad in reciprocal space and forms rods that extend along $\langle 110\rangle$.~\cite{Xu03:70,Xu04:69}  Several models have been proposed to explain this reciprocal space structure, including pancake-shaped regions in real space (Ref.~\onlinecite{Xu03:70,Welberry05:38,Welberry06:74}), polar domain walls oriented along $\langle 110 \rangle$ (Ref.~\onlinecite{Pasciak07:76}), correlations between chemically-ordered regions (Ref.~\onlinecite{Ganesh10:81}), Huang scattering (Ref.~\onlinecite{Vak05:7,Vak10:400}), and relatively isotropic displacements of the lead cations (Ref.~\onlinecite{Bosak11:xx}).  Obviously no consensus or satisfactory description currently exists of the real space structure of the short-range, polar correlations that give rise to the large and temperature-dependent elastic diffuse scattering cross section in these relaxor materials.
		
	The similarity of the reciprocal space geometries of the diffuse scattering in PMN and PZN demonstrates that they are structurally similar on short length scales.  However both compounds exhibit similar long-range ordered, average cubic crystal structures as well.  At high temperatures both systems possess cubic Pm$\overline{3}$m symmetry, while at low temperatures an average cubic unit cell is retained, at least insofar as neither compound undergoes a bulk structural phase transition as would be revealed by the splitting of a Bragg peak.  This low-temperature, average cubic crystal structure has been well established for PMN using both x-ray and neutron scattering techniques.~\cite{Bonneau89:24,Bonneau91:91,deMathan91:03}  However the case of PZN is controversial.  Variable energy x-ray diffraction studies designed to probe the bulk and near-surface regions of single crystal PZN observed a cubic phase in the bulk and a rhombohedral phase in the ``skin", which led to the discovery of an anomalous skin effect in relaxors.~\cite{Xu03:67}  This discovery was later challenged by other researchers who, based on neutron powder diffraction measurements, concluded that PZN exhibits uniform rhombohedral ground state structure instead.~\cite{Kisi05:17}  On this subject the weight of experimental evidence seems to favor an average cubic ground state structure for bulk PZN because neutron diffraction measurements on pure PZN are plagued by enormous extinction effects and thus effectively probe only the near-surface region, and because similar skin effects were subsequently reported by x-ray diffraction studies of single crystal PZN doped with 4.5\% PT and 8\% PT (PZN-$x$PT).~\cite{Xu04:84,Xu06:79}  Moreover, various diffraction studies of single crystal PMN doped with 10\% and 20\% PT (PMN-$x$PT) show that while the near-surface regions (probed by x-rays) are rhombohedrally distorted,~\cite{Dkhil01:65} the interior or bulk of these crystals (probed by neutrons) remains metrically cubic down to low temperatures, thus confirming the skin effect in this system.~\cite{Gehring04:16,Xu03:68}  Subsequent neutron-based strain experiments found a significant skin effect in large single crystals of PMN as well that has since been observed with x-rays,~\cite{Conlon04:70,Stock:unpub,Xu06:79,Pasciak11:226} and local regions of polarization have been directly imaged near the surface using piezoresponse force microscopy.~\cite{Shvartsman11:108}  Under the application of strong electric fields the bulk unit cell in PMN remains cubic, whereas the structure of the near-surface region distorts; however the intensity of the diffuse scattering arising from short-range, polar correlations can be suppressed by an external electric field, but only below the critical temperature $T_C \sim 210$\,K, which is indicative of a more ordered structure in the material.~\cite{Stock:unpub,Vak98:40}
	
	In view of these results we believe that the dielectric properties of both PMN and PZN can be described in terms of just two temperature scales:  a high-temperature scale $T_d$, below which static, short-range, polar correlations first appear, and a low-temperature scale $T_C$, below which long-range, ferroelectric correlations can be induced by cooling in the presence of a sufficiently strong electric field.  Each of these temperature scales is reflected in the neutron elastic diffuse scattering cross section in PMN when measured as a function of temperature and electric field,~\cite{Stock:unpub} as shown schematically in the upper panel of Fig.~\ref{diffuse_cte}.  The high-temperature scale $T_d$ is defined as the temperature at which the elastic diffuse scattering cross section becomes non-zero, and the low-temperature scale $T_C$ is defined as the temperature at which the elastic diffuse scattering can be suppressed by electric fields and ferroelectric domains are formed.  As shown by a recent high-resolution spin-echo study of the elastic diffuse scattering in PMN,~\cite{Stock10:81} the dynamics of relaxors also reflects these same two temperature scales.  The onset of diffuse scattering at $T_d$ was found to coincide with the appearance of static (at least on gigahertz timescales) polar nanoregions that coexist with dynamic, short-range, polar correlations.  $T_C$ was found to coincide with the temperature at which all short-range, polar correlations are static (within experimental resolution).~\cite{Stock10:81}  These two temperature scales are also clearly seen in measurements of the bulk polarization in La-doped PbZr$_{1-x}$Ti$_x$O$_3$ (PLZT) where $T_C$ marks the temperature at which the field-cooled and zero-field-cooled polarizations diverge.~\cite{Viehland92:46}  These two temperature scales can be understood in terms of random dipolar fields that are introduced through the disorder that is inherently present on the B-site of lead-based perovskite relaxors.~\cite{Stock04:69,Gvas05:17,Westphal92:68,Fisch03:67}  As a function of Ti content $x$, it has also been shown that $T_d$ and $T_C$ become equal at the morphotropic phase boundary (MPB), where the relaxor properties are suppressed in favor of those associated with a well-defined ferroelectric phase.~\cite{Cao08:78}

	While the static structure of relaxors displays a range of interesting properties, the dynamics associated with the relaxor transition have many unusual features as well.  The conventional picture of a (displacive) phase transition from a paraelectric state to a ferroelectric state is characterized by a long-wavelength ($q=0$) soft transverse optic (TO) phonon for which the frequency approaches a minimum at $T_C$, as happens in the well-known paraelectric-cubic to ferroelectric-tetragonal phase transition in PbTiO$_3$.~\cite{Shirane70:2,Dove:book}   The frequency of the soft mode $\Omega_{TO}$ is directly related to the dielectric constant $\epsilon$ determined from bulk measurements via the Lyddane-Sachs-Teller (LST) relationship, which states that $1 / \epsilon \propto (\hbar \Omega_{TO})^2$.  Therefore, neutron inelastic scattering measurements of the lattice dynamics provide an important tool with which to investigate ferroelectricity. In relaxors, the first pioneering work on the lattice dynamics of PMN was conducted by Naberezhnov \textit{et al.}~\cite{Nab99:11}  Experiments performed by Gehring \textit{et al.} and Wakimoto \textit{et al.} demonstrated the existence of a soft TO mode in PMN and tracked its temperature dependence.~\cite{Gehring01:87,Wakimoto02:65}   The soft mode in PMN reaches a minimum energy around 400\,K and then recovers linearly (i.\ e.\ $\Omega_{TO}^2 \propto |T - T_d|$) at lower temperatures.  As shown in Fig.~\ref{diffuse_cte}, 400\,K closely matches the onset temperature of the elastic (i.\ e.\ static) diffuse scattering cross section in PMN when measured on gigahertz timescales using neutron spin-echo and neutron backscattering methods.~\cite{Stock10:81}  Motivated by these results, Gehring \textit{et al.} proposed that the upper temperature scale $T_d$, more commonly known as the Burns temperature, should be normalized from the value of 620\,K determined from the measurements of the refractive index in PMN, to $420 \pm 20$\,K, where the soft TO mode reaches a minimum and static ferroelectric correlations (i.\ e.\ polar nanoregions) first appear.~\cite{Gehring09:79}
	
	An important anomalous feature found in these studies concerns the TO mode measured near the Brillouin zone center, which is unusually broad in energy compared to that in PbTiO$_3$ and even becomes overdamped, i.\ e.\ the half-width at half-maximum (HWHM) of the phonon energy lineshape $\Gamma$ exceeds the phonon energy $\hbar \Omega_{TO}$ over a range of temperatures and wavevectors; this indicates that some mechanism is present that shortens the lifetime of long-wavelength TO lattice fluctuations.  The term ``waterfall" was coined to describe this effect, and originally the broadening was believed to be the result of scattering of TO phonons from polar nanoregions.~\cite{Tsurumi94:33,Gehring00:84}  However, the same effect is seen in ferroelectric PMN-60\%PT, which exhibits no diffuse scattering and is not a relaxor.~\cite{Stock06:73}  This proves that the anomalous TO broadening cannot be associated with the relaxor phase but may be the result of the disorder introduced through the heterovalent nature of the B-site cations.  Another explanation for the broadening was expressed in terms of coupling between the TO and TA (transverse acoustic) modes as is observed in BaTiO$_3$;~\cite{Hlinka03:91} more recently this same idea has been used to explain the apparent ``waterfall" effect in PbTe.~\cite{Delaire11:10}  However a series of other papers have reported the existence of a second, quasi-optic mode at low-energies in PMN, which would then complicate the dynamics near the zone center, thus giving the appearance of a broadening or an extra neutron inelastic scattering cross section at low energies.~\cite{Vak02:66,Zein10:105}  This interpretation is particularly attractive as it could reconcile the apparent discrepancies in the dielectric response with the frequency of the low-energy TO mode expected from the LST relation described above as well as the measured deviations from the expected Curie-Weiss behavior.~\cite{Viehland92_2:46}  Still other groups claim that the extra spectral weight at low energies originates from quasi-elastic scattering from short-range, ferroelectric ordering.~\cite{Gvas05:17,Gvas04:49}
	
	It is quite clear given the many different aforementioned studies that the low-energy neutron scattering cross section is neither well understood nor well characterized in relaxors.  In this paper, we provide a detailed study of the Brillouin zone and temperature dependence of the transverse acoustic phonons in PMN over the entire Brillouin zone.  We observe highly anisotropic acoustic lattice fluctuations that reflect the existence of a coupling between the acoustic harmonic modes and the relaxational diffuse scattering.  We also provide measurements of the elastic constants obtained on the terahertz timescale and compare these with those made using other techniques and those of the canonical ferroelectric PbTiO$_3$.  Our results point to a reduced value of C$_{11}$ and C$_{12}$ in PMN compared to those in PbTiO$_3$.  Further, marked changes in the elastic anisotropy (2C$_{44}$/(C$_{11}$-C$_{12}$)) and differences in C$_{11}-$C$_{12}$ are reported here for PMN versus those in PbTiO$_3$.  We discuss these results in terms of the models mentioned earlier in order to develop a more complete description and understanding of the low-energy lattice fluctuations in PMN.

\section{EXPERIMENTAL DETAILS}

	All of the neutron scattering data presented in this paper were obtained using the C5 and N5 thermal-neutron, triple-axis spectrometers located at Chalk River Laboratories (Chalk River, Ontario, Canada), the SPINS cold-neutron, triple-axis and DCS time-of-flight spectrometers located at the National Institute of Standards and Technology (NIST) Center for Neutron Research (Gaithersburg, Maryland, USA).  Measurements made on C5 concentrated on the low-temperature phonons.  Inelastic data were measured using a variable vertical-focus, pyrolytic graphite (PG) monochromator and a flat PG analyzer.  The (002) Bragg reflections of the monochromator and analyzer crystals were used to analyze the incident and scattered neutron energies, respectively.  For elastic measurements, which characterize the nuclear Bragg peaks, the PG analyzer was replaced by a nearly perfect (mosaic $\sim 0.06^{\circ}$ FHWM), single-crystal, germanium (220) analyzer.  A PG filter was inserted into the scattered beam to remove higher order neutrons for all thermal neutron measurements.  A sapphire filter was placed before the monochromator to remove high-energy fast neutrons.  The inelastic data have been corrected for contamination of the incident beam monitor as described elsewhere (see Ref.~\onlinecite{Shirane:book} and the appendix of Ref.~\onlinecite{Stock04_2:69}).  The horizontal beam collimations were either 33$'$-33$'$-$S$-29$'$-72$'$ or 12$'$-33$'$-$S$-29$'$-72$'$ ($S$ = sample).  The final energy was fixed to either $E_f=14.6$\,meV or 14.8\,meV.  Data collected on N5 were taken using a flat PG monochromator and analyzer with a PG filter inserted into the scattered beam to remove higher order neutrons.  Horizontal beam collimations were set to 33$'$-26$'$-$S$-24$'$-open$'$ with a final energy of $E_f=14.6$\,meV.  These configurations provided an energy resolution at the elastic line of 0.95 $\pm$ 0.10 meV.  A cryofurnace was used to control the sample temperature on both C5 and N5, and helium exchange gas was inserted into the sample space to ensure good thermal contact and equilibrium at all temperatures.  A thermometer was also placed on the sample stick near the sample position in order to verify and monitor the sample temperature.

	Data on SPINS were obtained using a variable vertical-focus PG monochromator.  A flat PG analyzer was inserted into the scattered beam in series with a beryllium filter to remove higher order neutrons, and the final energy was fixed to $E_f=4.5$\,meV.  The collimations were set to guide-80$'$-$S$-80$'$-open for inelastic measurements and guide-10$'$-$S$-10$'$-open for studies of the nuclear Bragg peaks.  The energy resolution for the inelastic configuration was 0.28 $\pm$ 0.05 meV.  The sample was placed in a high-temperature displex so that temperatures from 50\,K to 600\,K could be reached.

	Data collected on DCS were taken using a fixed incident energy of 6.0\,meV, and the time between pulses at the sample was set to 9\,ms, which effectively eliminates frame overlap.  Time-of-flight spectra were collected for $\sim 45$ different crystal orientations spaced $0.5^{\circ}$ apart at each temperature using the 325 detectors with an angular coverage from $5-140^{\circ}$.  More details about the DCS instrument can be found elsewhere.~\cite{Copley00:283} The energy resolution at the elastic line was measured to be 0.28 $\pm$ 0.03 meV.  The sample was mounted in a high-temperature, closed-cycle $^4$He-refrigerator such that temperatures between 50\,K to 600\,K could be reached.

	The data presented in this paper were obtained from two different single crystals of PMN, which were aligned in either the (HK0) or (HHL) scattering planes.  The room-temperature, cubic lattice constant is $a = 4.04$\AA, so one reciprocal lattice unit (rlu) equals $2\pi/a = 1.56$\,\AA$^{-1}$.  The crystal labeled S-I is 9\,cc in volume, and the second crystal labeled S-II is 5\,cc in volume.  We found no evidence of a structural distortion in either sample.  Crystal S-I contains a very small amount of PbTiO$_3$ that was intentionally added during sample preparation to facilitate the growth of large single crystals; the sample preparation method has been described previously.~\cite{Luo00:39}  Crystal S-I is the same sample as that used in a previous study of the low-energy phonons (see Ref.~\onlinecite{Stock05:74}).  We emphasize that the phonons we measured are identical across all samples studied.  We compare our measurements on these PMN crystals to those on PMN-60\%PT, which is \textit{not} a relaxor and which undergoes a well-defined, cubic-to-tetragonal phase transition just below 550\,K.~\cite{Stock06:73}  We also compare our results to published and unpublished measurements made on a smaller PMN single crystal that was studied in a series of earlier publications.~\cite{Gehring09:79,Wakimoto02:65,Waki02:66}

\section{ELASTIC SCATTERING: ANISOTROPIC DIFFUSE SCATTERING AND SHORT-RANGE, POLAR CORRELATIONS}

	In this section, we discuss the temperature dependence of the diffuse scattering cross section as well as its dependence on wavevector when projected onto the (HHL) scattering plane.  Our results confirm previous experimental findings that the diffuse scattering is highly anisotropic in reciprocal space.  We also discuss how the two temperature scales defined in the Introduction are manifested in the diffuse scattering data and the consequences for the polar and ferroelectric correlations.

	The temperature and wavevector dependences of the neutron diffuse scattering cross section in PMN have been reported in several studies in which measurements were made mainly in the (HK0) scattering plane.~\cite{Hiraka04:70,Xu04:69}  These studies found that the distribution of diffuse scattering intensity in reciprocal space is composed of rods that are elongated along $\langle 110 \rangle$, which agrees with the results of the three-dimensional mapping study conducted with high-energy x-rays on PZN doped with PbTiO$_3$.~\cite{Xu04:70}  The reciprocal space anisotropy revealed by these measurements has been interpreted in terms of an underlying two-dimensional, real space structure in which the long axis of the diffuse rod reflects short-range, polar correlations while the shorter, perpendicular axes reflect longer-range correlations.  Fig.~\ref{diffuse_summary} shows data obtained on the cold-neutron, triple-axis spectrometer SPINS that illustrate how the neutron diffuse scattering cross section in PMN looks when projected onto the (HHL) scattering plane.  Panel $(a)$ shows constant-intensity contours measured near $\vec{Q}=(0,0,1)$ while panel $(b)$ shows contours measured near $\vec{Q}=(1,1,0)$.  In both cases a butterfly-shaped pattern is seen that appears to be slightly elongated roughly along $\langle 111 \rangle$.  This is consistent with neutron measurements performed in the (HK0) scattering plane as well as those from the three-dimensional x-ray mapping study, which observed rods extending along $\langle 110 \rangle$.  The slight elongation along $\langle 111 \rangle$ results from rods that protrude out of the scattering plane along [1$\overline{1}$0], but which nonetheless contribute to the total diffuse scattering pattern because of the non-zero vertical (i.\ e.\ out-of-plane) instrumental $Q$-resolution.  These results demonstrate that the diffuse scattering cross section is highly anisotropic in reciprocal space.  In particular, Fig.~\ref{diffuse_summary} $(a)$ shows that the diffuse scattering measured near $\vec{Q}=(0,0,1)$ is slightly narrower along [001] than it is along [110] while the opposite is true for the diffuse scattering measured near $\vec{Q}=(1,1,0)$ (Fig.~\ref{diffuse_summary} $(b)$).  As mentioned in the Introduction, this reciprocal space anisotropy has been interpreted in terms of a correspondingly anisotropic real space structure composed of local, two-dimensional correlations.
	
\begin{figure}[t]
\includegraphics[width=80mm]{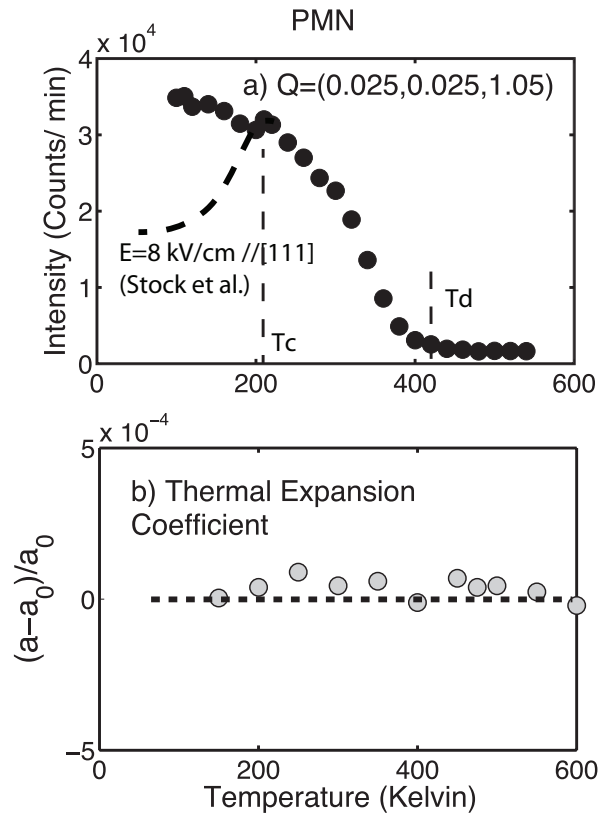}
\caption{$a)$ Elastic diffuse scattering intensity measured in zero field at $\vec{Q}=(0.025,0.025,1.05)$ plotted versus temperature. The data were measured on SPINS using a final energy $E_f=4.5$\,meV.  The sample (S-II) was aligned in the (HHL) scattering plane. The dashed line is a schematic representation of the field-cooled data presented in Ref.~\onlinecite{Stock:unpub}.  $b)$ The thermal expansion coefficient is plotted as a function of temperature, illustrating no observable structural distortion to a sensitivity of $\alpha \sim 10^{-4}$. The data were taken on C5 with $E_f=14.6$\,meV using a perfect germanium analyzer for improved $Q$-resolution. The sample (S-II) was aligned in the (HK0) scattering plane.}
\label{diffuse_cte}
\end{figure}

	The temperature dependence of the diffuse scattering cross section measured at $\vec{Q}=(0.025,0.025,1.05)$ is summarized in Fig.~\ref{diffuse_cte} $(a)$; this value of $\vec{Q}$ corresponds to the peak in the diffuse scattering intensity when the wavevector is scanned along (H,H,1.05).  These measurements were performed using cold neutrons on SPINS with a final neutron energy $E_f=4.5$\,meV, which provides an excellent energy resolution (FWHM) of $2\delta E \sim 0.23$\,meV.  The data show a well-defined onset of the diffuse scattering intensity at $T_d \sim 420 \pm 20 $\,K that matches the onset measured using far better energy resolution ($2\delta E \sim 1$\,$\mu$eV) provided by neutron backscattering techniques.~\cite{Gehring09:79} The issue of energy resolution is extremely relevant here because when the diffuse scattering is measured using a technique (e.\ g.\ x-rays) that provides a substantially coarser energy resolution the resulting intensity will necessarily include an integration over the low-frequency (quasielastic) component of the diffuse scattering cross section, and this will lead to an artificially higher (incorrect) onset temperature.~\cite{Hiraka04:70}  Given the invariance of the onset temperature of the elastic diffuse scattering when changing the instrumental energy resolution by more than two orders of magnitude, we conclude that $T_d \sim 420 \pm 20 $\,K represents an intrinsic and physically-meaningful temperature scale in PMN.  The same onset temperature was obtained for several different values of $\vec{Q}$ on different backscattering and neutron spin echo spectrometers; hence $T_d$ it is not wavevector dependent.  We further note that this temperature matches the Curie-Weiss temperature derived from linear fits of the inverse dielectric susceptibility measured at a frequency of 100\,kHz;~\cite{Viehland92_2:46} it also matches the temperature at which the soft TO mode energy in PMN reaches a minimum value.~\cite{Wakimoto02:65,Stock05:74,Stock10:81}

	Cooling PMN below $T_d$ does not induce a well-defined phase transition to a long-range ordered structure; this is evident from Fig.~\ref{diffuse_cte} $b)$, where the thermal expansion coefficient of the S-II PMN crystal is plotted from just below 200\,K to 600\,K.   Near a conventional ferroelectric phase transition, an anomaly is observed in the temperature dependence of the lattice constant like that seen at the ferroelectric transition of PbTiO$_3$;~\cite{Shirane51:6} however no such anomaly is seen in PMN.  In the relaxor PZN, the coefficient of thermal expansion exhibits an anomaly of $\alpha \sim 5 \times 10^{-4}$ at $T_C=400$\,K in the near-surface region.~\cite{Xu04_2:70}  Efforts to locate a  structural distortion in PMN were carried out using the C5 thermal-neutron and the SPINS cold-neutron, triple-axis spectrometers.  A configuration that provided excellent wavevector resolution was obtained on C5 by closely matching the $d$-spacing of the sample and that of a nearly perfect Germanium (220) analyzer crystal.  This configuration and the resulting resolution function are reviewed elsewhere.~\cite{Xu03:xx}  The data shown in Fig.~\ref{diffuse_cte} $b)$ provide no sign of a change in the dimensions of the unit cell; instead they imply an Invar-like behavior of the lattice constant up to 600\,K.  Therefore, even though $T_d$ represents a meaningful temperature scale in PMN that is associated with the onset of static, short-range, polar order and the softening of the optic mode, it is not associated with a structural phase transition to long-range, polar order.
	
	In the bulk of single crystal PMN and PZN there is no evidence for a well-defined structural transition.  Several powder diffraction studies of pure PMN have seen subtle indications of a structural distortion at temperatures below 200\,K, but these are well-modeled by a disordered lattice with local atomic shifts.~\cite{Bonneau89:24,Bonneau91:91,Dkhil01:65}  The temperature of 200\,K corresponds closely to that at which a sufficiently strong electric field applied along [111] is able to suppress the diffuse scattering cross section in PMN (see Fig. \ref{diffuse_cte} $(a)$) while simultaneously inducing a structural distortion in the near-surface region.~\cite{Stock:unpub}  This temperature scale is denoted as $T_C$ in Fig.~\ref{diffuse_cte} $(a)$.  Other studies of single crystal PMN and PMN substituted with 10\% PbTiO$_3$ (PMN-10\%PT) using neutron scattering have found no evidence of any anomaly in the lattice constant at low temperatures, but they do observe a significant change above $\sim 350$\,K (Ref.~\onlinecite{Gehring09:79,Gehring04:16}).  These findings are consistent with x-ray diffraction studies of ceramic samples of PMN-10\%PT,~\cite{King04:31} but they differ from the result reported here in Fig.~\ref{diffuse_cte} $b)$.  Strain-scanning experiments performed on single crystals using neutrons have found evidence for a significant near-surface effect in PMN and even large changes of the coefficient of thermal expansion as a function of depth in the crystal.~\cite{Conlon04:70,Xu06:79}  We therefore conclude that no well-defined structural transition occurs at $T_d \sim 420$\,K, and that the apparent thermal expansion at higher temperatures is sample dependent and likely linked to the effects of a skin layer in these materials.  However, the Invar-like response in the bulk at temperatures below $\sim 350$\,K is consistent across a series of PMN samples and all known studies.
	
	Based on these data and those presented in previous studies, we conclude that the diffuse scattering cross section is highly anisotropic in reciprocal space.  This conclusion is consistent with previous measurements that describe the diffuse scattering as being composed of rods oriented along $\langle 110 \rangle$.~\cite{Xu04:70}  Elastic diffuse scattering first appears below $T_d \sim 420$\,K,~\cite{Gehring09:79} a temperature scale that does not reflect a phase transition to a long-range ordered structure, but rather to the onset of static, short-range, polar correlations.  A longer-range ferroelectric phase can be induced by cooling in the presence of a sufficiently strong electric field, but only below $T_C \sim 210$\,K.~\cite{Stock:unpub}

\section{INELASTIC SCATTERING:  ANISOTROPIC TRANSVERSE ACOUSTIC PHONON DAMPING }

\begin{figure}[t]
\includegraphics[width=85mm]{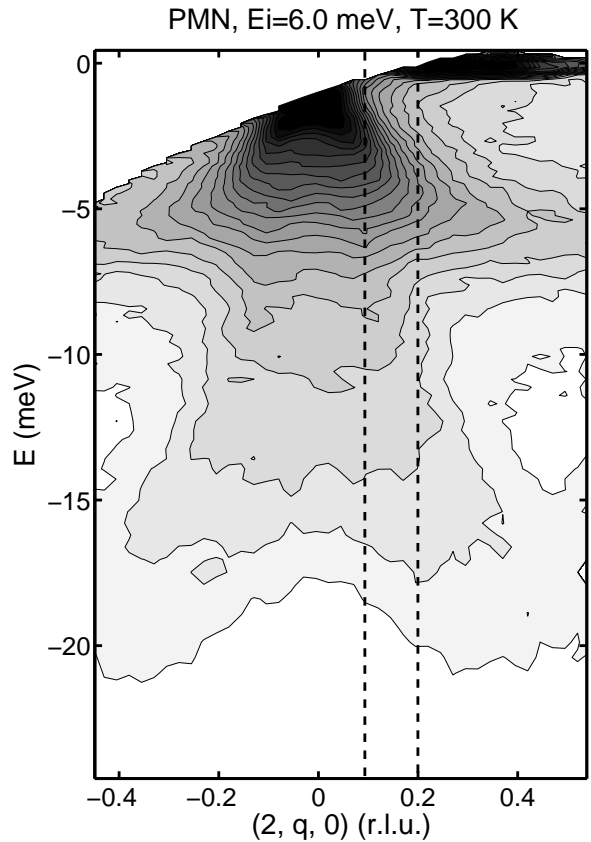}
\caption{DCS neutron scattering data measured on PMN sample S-II in the (HK0) scattering plane near (200) and along [010] at $T=300$\,K.  The constant inelastic scattering intensity contours reveal well-defined TA$_1$ and TO modes near the zone boundary at $(2,\pm0.5,0)$, but overdamped TO modes near the zone center.  The vertical dashed lines indicate where constant-$\vec{Q}$ scans were made; these are discussed in the next figure.}
\label{waterfall}
\end{figure}

So far we have reviewed the wavevector and temperature dependence of the neutron elastic diffuse scattering cross section in PMN, which reflects the presence highly anisotropic, static short-range, polar correlations that develop below $T_d \sim 420$\,K.  We now present extensive neutron inelastic scattering measurements of the transverse acoustic (TA) phonons that show how the anisotropy of these short-range, polar correlations is manifested in the damping of acoustic lattice fluctuations.  To this end, we discuss the TA phonon scattering cross section measured in three different Brillouin zones and the temperature dependence observed in each case.

We chose to make our neutron scattering measurements in the (HK0) scattering plane because they are then sensitive to two different TA phonons, denoted by TA$_1$ and TA$_2$.  This sensitivity is governed by the neutron scattering phonon cross section, which is proportional to $(\vec{Q} \cdot \vec{\xi})^2$, where $\vec{Q}$ is the total momentum transfer, or scattering vector, and $\vec{\xi}$ is the eigenvector of the phonon.  Measurements made at reduced wavevectors $\vec{q}$ transverse to (100) or (200), e.\ g.\ for $\vec{Q} = (1,q,0)$, are sensitive to TA$_1$ acoustic phonons, which propagate along [010] and are polarized along [100].  The limiting slope of the phonon dispersion associated with this mode as $q \rightarrow 0$ is proportional to the C$_{44}$ elastic constant.  Measurements made at $\vec{q}$ transverse to (110) or (220), e.\ g.\ for $\vec{Q} = (1+q,1-q,0)$, are sensitive to TA$_2$ phonons, which propagate along [1$\overline{1}$0] and are polarized along [110].  The limiting slope of this phonon dispersion is proportional to ${ 1\over2}($C$_{11}-$C$_{12})$.  A detailed list of the elastic constant dependence based on scan direction can be found in Ref.~\onlinecite{Dove:book} and Ref.~\onlinecite{Noda89:40}.

In this section we will first consider the TA$_1$ and TA$_2$ acoustic phonons measured in the (200) and (220) Brillouin zones as these zones exhibit relatively weak neutron diffuse scattering cross sections but strong acoustic and optic phonon scattering cross sections.  We will then examine the TA$_1$ and TA$_2$ acoustic phonons measured in the (100) and (110) Brillouin zones where strong diffuse scattering cross sections are present.  This fact has been demonstrated quantitatively in PMN at room temperature by Vakhrushev \textit{et al}., who found that the ratio of the diffuse scattering intensities measured in the (110) and (220) Brillouin zones was $I_d(110)/I_d(220) \textgreater 63/4 \sim 16$ and that between the (300) and (200) Brillouin zones was $I_d(300)/I_d(200) \textgreater 87/4 \sim 22$.~\cite{Vak95:37}  These ratios should be viewed as lower bounds because the experimental determinations of the weak diffuse scattering cross sections in the (200) and (220) zones were limited by experimental uncertainties.  We mention that, although our data cover a wide range of $q$ that spans each Brillouin zone, our discussion will not include the highly damped, soft optic phonons that are observed at various zone boundaries, because they have already been studied in detail in Ref.~\onlinecite{Swainson09:79}.

\subsection{$\vec{Q}=(2,0,0)$ - TA$_1$ acoustic phonons in the presence of weak diffuse scattering}

\begin{figure}[t]
\includegraphics[width=90mm]{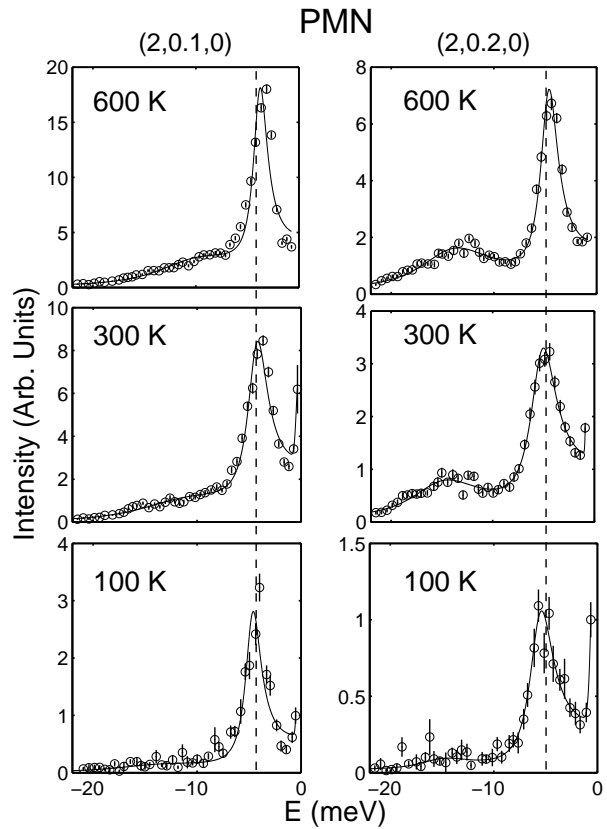}
\caption{Temperature dependence of TA$_1$ and TO phonons measured at two different wavevectors on PMN sample S-II in the (HK0) scattering plane using the DCS spectrometer.  The constant-$\vec{Q}$ scans integrate over the regions $(2 \pm 0.025, 0.10 \pm 0.02, 0)$ and $(2 \pm 0.025, 0.20 \pm 0.025, 0)$. The vertical dashed lines indicate the average position in energy of the TA$_1$ phonon.}
\label{T1_Q200}
\end{figure}

While the (200) neutron diffuse scattering cross section is non-zero, as pointed out in Ref.~\onlinecite{Gehring09:79}, it is more than one order of magnitude smaller than those in the (100), (110), and (300) Brillouin zones, which is consistent with the structure factor calculations discussed in Ref.~\onlinecite{Hirota02:65} and Ref.~\onlinecite{Hiraka04:70}.  We therefore begin our discussion of the phonons in PMN in the (200) zone, where the acoustic and optic phonon scattering cross sections are strong and the diffuse scattering cross section is weak.

Fig.~\ref{waterfall} shows a contour map of the inelastic scattering intensity measured at 300\,K along [010] that spans the entire (200) Brillouin zone and covers the energy range from -20\,meV to 0\,meV.  The negative energy scale merely indicates that these data were obtained using a neutron energy gain configuration in which energy is transferred from the lattice to the neutron during the scattering process; this configuration was chosen because it provides a broader detector coverage of momentum-energy phase space.  At this temperature, well-defined TA$_1$ and TO phonon branches are seen at large $q$.  The TA$_1$ mode reaches a maximum energy of $\sim 6$\,meV at the X-point zone boundary ($\vec{Q}=(2,\pm0.5,0)$), whereas the TO mode is observed at higher energies and extends to $\sim 20$\,meV.  However for wavevectors approaching the zone center ($q=0$) the TO mode becomes overdamped while the TA$_1$ remains well-defined.  This behavior has been termed the ``waterfall" effect and was discussed in the Introduction.~\cite{Gehring00:84}  Previous studies have shown that the TO phonon broadens and softens near the zone center as a function of temperature, reaching a minimum energy at $T_d \sim 420$\,K.~\cite{Gehring01:87,Wakimoto02:65}  At temperatures below $T_d$ the TO phonon hardens and the linewidth narrows.

This effect is more readily seen in Fig.~\ref{T1_Q200}, which displays constant-$\vec{Q}$ scans measured at $q=0.1$\,rlu and $q=0.2$\,rlu at 600\,K, 300\,K, and 100\,K; these two constant-$\vec{Q}$ scans correspond to the two vertical dashed lines shown in Fig.~\ref{waterfall}.  All of the data were fit to two uncoupled harmonic oscillators in order to describe the TO mode and the TA$_1$ mode; the excellent fits are indicated by the solid curves.  For both small ($q=0.1$\,rlu) and large ($q=0.2$\,rlu) wavevectors, and at all temperatures, a strong, sharp, TA$_1$ mode is observed.  By contrast, the TO mode is only well-defined at large wavevector; at $q=0.1$\,rlu the TO mode is overdamped due to the waterfall effect.  On cooling from 600\,K the intensity of both modes decreases in line with expectations based on the Bose factor.   A detailed analysis of the temperature dependence of the structure factors and intensities of the TA modes in PMN has already been published.~\cite{Stock05:74}  On the basis of that study it was concluded that the TA-TO mode coupling in PMN is minimal and does not give rise to the waterfall effect.  The lack of TA-TO mode coupling is demonstrated by the vertical lines in Fig.~\ref{T1_Q200}, which show that the TA$_1$ phonon energy does not shift with temperature despite the softening and hardening of the optic mode.  If strong TA-TO mode coupling were present in PMN, as it is in KTaO$_3$ (Ref.~\onlinecite{Axe70:1}), then a large change in the acoustic mode energy position should occur with temperature that mimics the change in the TO mode frequency.  We observe no such change in our experiments.  We note that this conclusion differs from those of some other studies,~\cite{Hlinka03:91,Vak02:66} and we will discuss these differences at the end of this paper.

Independent of the issue of mode coupling, Fig.~\ref{T1_Q200} demonstrates that the TA$_1$ mode measured near (200) is underdamped (i.\ e.\ $\Gamma \textless \hbar \Omega_0$) at all wavevectors and temperatures studied, and that it does not shift significantly in energy with temperature.  As demonstrated in Ref.~\onlinecite{Stock05:74}, the TA phonon intensities are in reasonable agreement with harmonic theory.  We note that the TA$_2$ acoustic phonons measured near (220) also exhibit little to no shift in energy with temperature and are also underdamped at all temperatures studied, i.\ e.\ between 100\,K and 600\,K.  In both the (200) and (220) Brillouin zones the diffuse scattering cross section is small in comparison to those in other zones such as (100) and (110).  We now move on to discuss the TA phonon cross section in these zones, where the diffuse scattering cross section is very strong.

\subsection{$\vec{Q}=(1,1,0)$ - TA$_2$ acoustic phonons in the presence of strong diffuse scattering}

\begin{figure}[t]
\includegraphics[width=80mm]{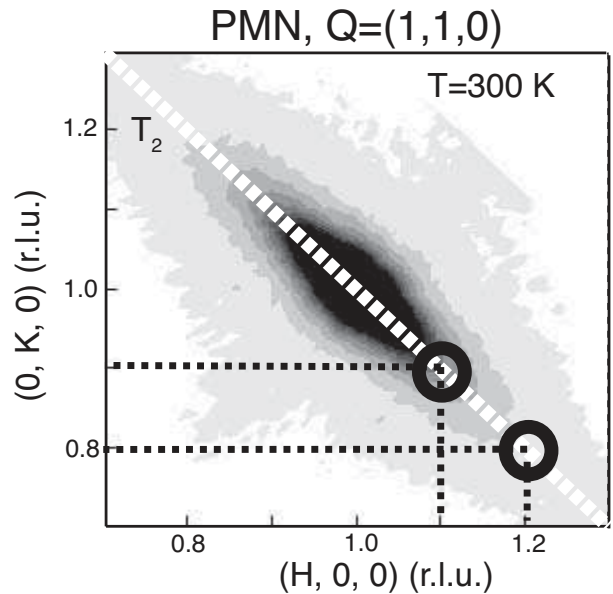}
\caption{Elastic diffuse scattering intensity contours measured near (110) on SPINS ($E_f=4.5$\,meV) in the (HK0) scattering plane.  The white dashed line represents the direction in reciprocal space where the TA$_2$ and TO phonons were measured.  The open black circles represent typical reciprocal lattice points where constant-$Q$ scans were measured.}
\label{summary_110}
\end{figure}

The white, dashed line in Fig.~\ref{summary_110} represents the set of reciprocal lattice points where constant-$\vec{Q}$ scans measured in the (110) Brillouin zone are sensitive to TA$_2$ phonons; as explained earlier, constant-$\vec{Q}$ scans directly probe TA$_2$ phonons when measured along a direction $\vec{q}$ transverse to $\vec{Q}=(1,1,0)$.  This figure also illustrates the relationship between these reciprocal lattice points and the elastic diffuse scattering cross section measured at 300\,K.   The diffuse scattering intensity contours were measured by Hiraka \textit{et al.} and are taken from Ref.~\onlinecite{Hiraka04:70}.  The white, dashed line also corresponds to a ridge along which the diffuse scattering cross section is a maximum compared to that along all other parallel lines.  Hence constant-$\vec{Q}$ scans measured at points on this dashed line will reflect TA$_2$ phonons in the presence of maximum diffuse scattering intensity.

\begin{figure}[t]
\includegraphics[width=75mm]{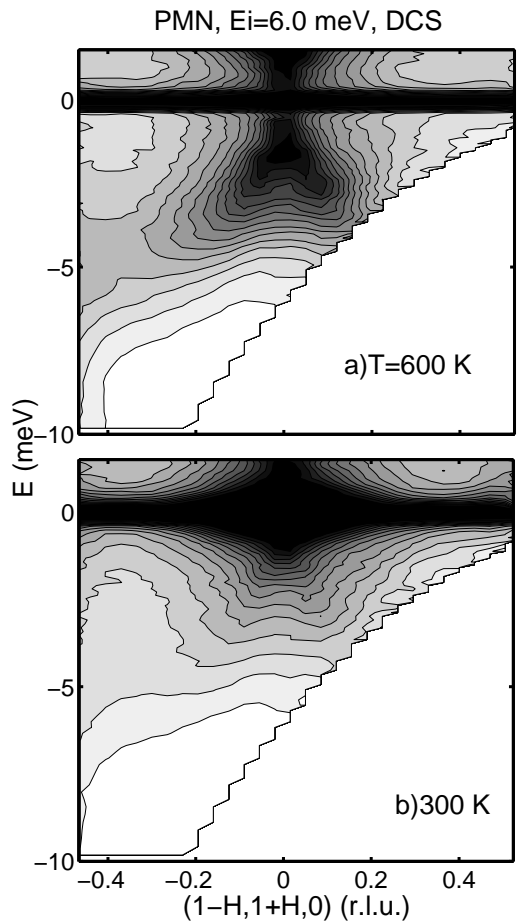}
\caption{Constant inelastic scattering intensity contours measured on DCS in the (110) Brillouin zone are shown at (a) 600\,K and (b) 300\,K after integrating over $(1 \pm 0.1, 1 \pm 0.1,0)$.  Underdamped, dispersive TA$_2$ phonons at 600\,K become heavily damped at all $q$ at 300\,K except for $q \sim 0$.  The TO mode is not visible in this zone due to a weak structure factor.  The wide, black streak at $E=0$\,meV is where the intensity is dominated by the incoherent, elastic scattering cross section from the sample and the aluminium sample mount.  These data were obtained on the S-II PMN crystal aligned in the (HK0) scattering plane.}
\label{dispersion_110}
\end{figure}

Reciprocal space maps spanning the (110) Brillouin zone show the TA$_2$ phonon dispersion along $[1\overline{1}0]$ at 600\,K and 300\,K in Fig.~\ref{dispersion_110}.  As mentioned before, because of a better detector coverage of momentum-energy phase space, our data were taken on the neutron energy gain side, which is indicated by the negative value of the energy transfers.  The constant inelastic scattering intensity contours in panel $a)$ provide evidence of a distinct and well-defined TA$_2$ phonon branch at 600\,K.  But the TO phonon branch that was seen in the (200) Brillouin zone is not visible here because of a very weak structure factor.  As in the (200) zone, the TA$_2$ mode reaches a maximum energy of $\sim 6$\,meV at the zone boundary and appears to be well-defined (underdamped) throughout most of the Brillouin zone.  The same data are shown after cooling to 300\,K in panel $b)$.  At this temperature the contours reveal a TA$_2$ spectrum that is very different from that at 600\,K because the TA$_2$ lineshape is now overdamped throughout most of the Brillouin zone.

\begin{figure}[t]
\includegraphics[width=80mm]{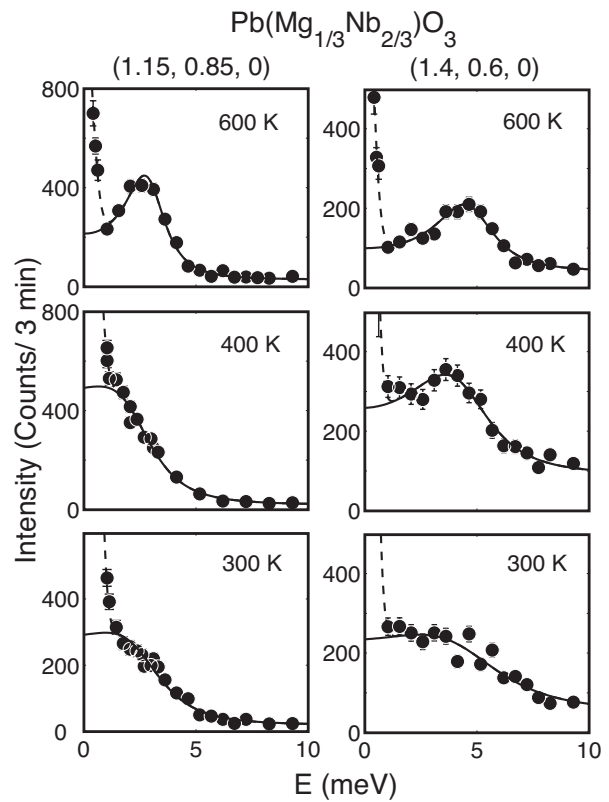}
\caption{Constant-$\vec{Q}$ scans measured at $\vec{Q}=(1.15,0.85,0)$ and (1.4,0.6,0) at temperatures ranging from 300-600\,K.  The solid lines are fits to a damped harmonic oscillator as described in the text.  The data were taken with the C5 thermal-neutron, triple-axis spectrometer located at Chalk River using $E_f=14.6$\,meV.  The PMN S-I sample was aligned in the (HK0) scattering plane.}
\label{T2_highT}
\end{figure}

To better understand the temperature dependence of the TA$_2$ phonon linewidth $\Gamma$ in the (110) Brillouin zone, we measured constant-$\vec{Q}$ scans similar to those shown in Fig.~\ref{T1_Q200}.  These scans are presented in Fig.~\ref{T2_highT}, which summarizes the response of the TA$_2$ acoustic phonon at temperatures well above, near, and below $T_d$, where the elastic diffuse scattering first appears.  The solid lines in Fig.~\ref{T2_highT} are fits to a harmonic oscillator lineshape that obeys the principle of detailed balance.~\cite{Shirane:book}  The dashed lines correspond to a Gaussian function of energy centered at $E=0$ and describe the elastic scattering cross section, which is a combination of diffuse scattering and incoherent scattering from the sample and sample mount.  Details of the lineshape and the fitting have presented previously in Ref.~\onlinecite{Stock05:74}.  At 600\,K underdamped TA$_2$ phonons are observed at $\vec{Q}=(1.15, 0.85,0)$ and $(1.4,0.6,0)$, which are consistent with the contours shown in Fig.~\ref{dispersion_110}.  However after cooling to 400\,K the TA$_2$ mode at $\vec{Q}=(1.15,0.85,0)$ becomes overdamped and resembles a ``quasi elastic" lineshape in that it overlaps with the elastic line at $E=0$\,meV.  A similar lineshape is observed at $\vec{Q}=(1.4, 0.6,0)$, which lies very near the M-point zone boundary.  Finally, at 300\,K, which is above $T_C$ but below $T_d$ where the soft TO mode achieves a minimum in energy, the TA$_2$ mode at both wavevectors is again described by an overdamped lineshape.

\begin{figure}[t]
\includegraphics[width=80mm]{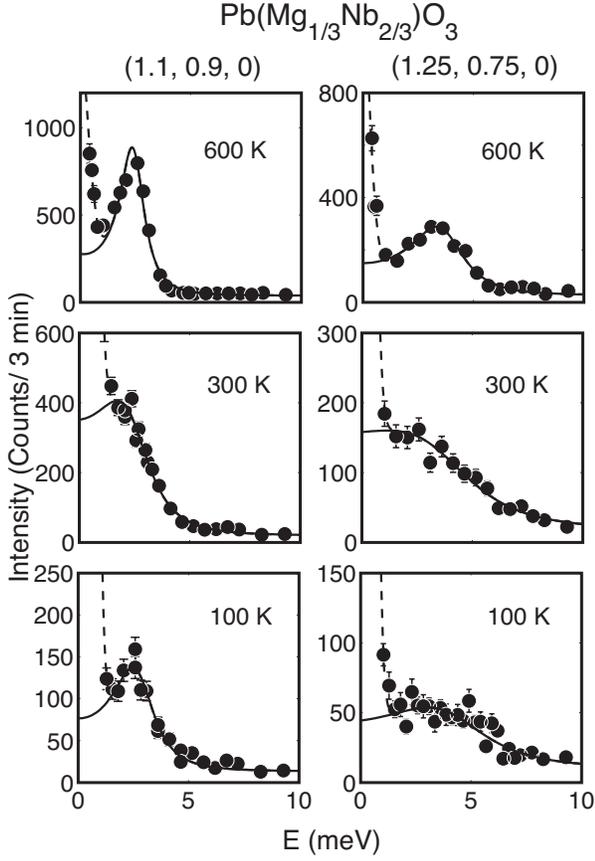}
\caption{Constant-$\vec{Q}$ scans measured at $\vec{Q}=(1.1,0.9,0)$ and $(1.25,0.75,0)$ at 600\,K, 300\,K, and 100\,K.  The solid lines represent fits to a damped harmonic oscillator as described in the text.  The data were taken with the C5 thermal-neutron, triple-axis spectrometer (Chalk River) using $E_f=14.6$\,meV.  The PMN S-I sample was aligned in the (HK0) scattering plane.}
\label{T2_allT}
\end{figure}

Figure~\ref{T2_allT} extends our study of the temperature dependence of the TA$_2$ linewidth to temperatures below T$_c \sim 200$\,K and wavevectors $\vec{Q}=(1.1,0.9,0)$ and $(1.25,0.75,0)$.  Both wavevectors exhibit an underdamped TA$_2$ lineshape at 600\,K and an overdamped lineshape at 300\,K; however at 100\,K, i.\ e.\ below $T_C$, the TA$_2$ phonon recovers an underdamped lineshape.  This low-temperature recovery is more evident at low wavevector: at $\vec{Q}=(1.1,0.9,0)$ the TA$_2$ phonon lineshape is quite sharp and the energy has hardened to a value comparable with that measured at 600\,K; by contrast the lineshape at $\vec{Q}$=(1.25,0.75,0) displays a more subtle recovery inasmuch as it still resembles an overdamped form.  We point out that at 100\,K (see Fig.~\ref{diffuse_cte}) the elastic diffuse scattering cross section is maximum.

These results provide overwhelming evidence of strong coupling between the TA$_2$ mode and the diffuse scattering centered at $E=0$.  The fact that strong, temperature-dependent shifts in the TA$_2$ phonon energy and changes in the TA$_2$ linewidth are observed in the (110) zone, where the TO phonon cross section is weak, but not in the (220) zone, where the TO phonon cross section is strong, proves that these effects are \textit{not} due to coupling between acoustic and optic modes.  Instead, these effects correlate directly with the strength and temperature dependence of the diffuse scattering cross section.  This is further corroborated by measurements of the TA$_{2}$ mode in PZN-4.5PT which demonstrated a recovery in the acoustic mode when an electric field suppressed the diffuse scattering cross section.~\cite{Xu08:7}    It has been shown in a series of studies describing the central peak in SrTiO$_3$ (Ref.~\onlinecite{Shapiro72:16}), and in the appendix of Ref.~\onlinecite{Stock05:74}, that a dynamic component of the diffuse scattering can couple to a low-energy, harmonic mode to produce an overdamped lineshape.  Theoretical studies have also discussed this possibility.~\cite{Halperin76:14}  Recent studies of PMN-30\%PT have applied this model to try to understand the heavily-damped lineshapes observed for the acoustic mode.~\cite{Matsuura11:80}  Studies of the diffuse scattering using neutron spin-echo (NSE) techniques, which provide a considerable dynamic range in Fourier time, have documented the presence of such a dynamic component over a wide range in temperatures, which becomes static below $T_C \sim 210$\,K.~\cite{Stock10:81}  Broad-band dielectric measurements suggest that the dynamics may even persist to much lower frequencies than are probed by the NSE technique.~\cite{Bovtun06:26,Kamba05:17}  To further support the conjecture that the TA phonon damping documented here is the result of a coupling between the acoustic and diffuse scattering cross sections, we now present measurements of the TA$_1$ phonons in the (100) Brillouin zone.

\subsection{$\vec{Q}$=(1,0,0) - TA$_1$ acoustic phonons in the presence of strong diffuse scattering}

\begin{figure}[t]
\includegraphics[width=80mm]{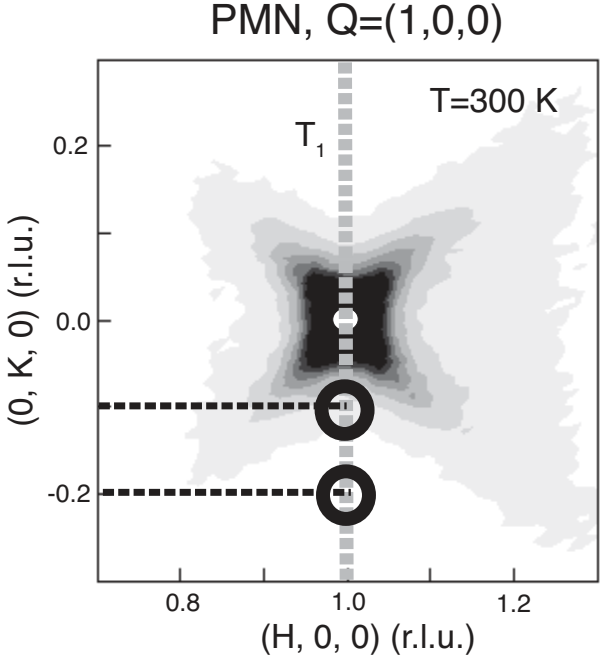}
\caption{Elastic diffuse scattering intensity contours measured near (100) on SPINS ($E_f=4.5$\,meV) in the (HK0) scattering plane.  The grey, dashed line represents the direction in reciprocal space where the TA$_1$ and TO phonons were measured.  The open black circles represent typical reciprocal lattice points where constant-$Q$ scans were measured.}
\label{summary_100}
\end{figure}

The grey, dashed line in Fig.~\ref{summary_100} represents the set of reciprocal lattice points where constant-$\vec{Q}$ scans measured in the (100) Brillouin zone are sensitive to TA$_1$ phonons; as explained earlier, constant-$\vec{Q}$ scans directly probe TA$_1$ phonons when measured along a direction $\vec{q}$ transverse to $\vec{Q}=(1,0,0)$.  This figure also illustrates the relationship between these reciprocal lattice points and the elastic diffuse scattering cross section measured at 300\,K.   These diffuse scattering intensity contours were also measured by Hiraka \textit{et al.} and are again taken from Ref.~\onlinecite{Hiraka04:70}.  However, unlike the white, dashed line shown in Fig.~\ref{summary_110}, the grey, dashed line here does \textit{not} correspond to a ridge along which the diffuse scattering cross section is a maximum.  Thus the reciprocal lattice points used to study the TA$_1$ phonon mode in the (100) Brillouin zone are \textit{not} located on ridges of maximum diffuse scattering, as was the case for the TA$_2$ phonons in the (110) Brillouin zone; instead they are located between them where the diffuse scattering cross section is nonetheless strong.

We briefly note here that Fig.~\ref{summary_100} shows diffuse scattering intensity contours measured in the (HK0) scattering plane whereas the those in Fig.~\ref{diffuse_summary} were measured in the (HHL) plane.  It can be readily seen that the elastic diffuse scattering cross-section is composed of rods that are extended along the [1$\overline{1}$0] and [$\overline{1}$10] directions.

\begin{figure}[t]
\includegraphics[width=92mm]{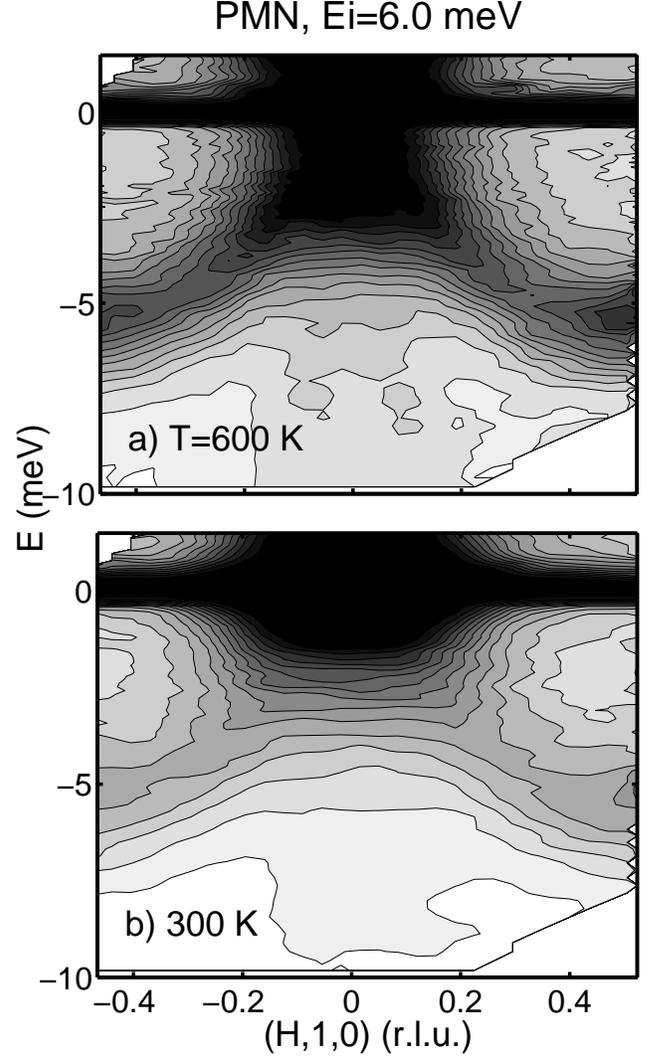}
\caption{Constant inelastic scattering intensity contours measured on DCS in the (100) Brillouin zone are shown at (a) 600\,K and (b) 300\,K after integrating over $(0, 1 \pm 0.15,0)$.  Underdamped, dispersive TA$_1$ phonons at 600\,K remain well-defined near the zone boundary at 300 K, but become heavily damped for intermediate wavevectors.  The wide, black streak at $E=0$ is where the intensity is dominated by the incoherent, elastic scattering cross section from the sample and the aluminium sample mount.  These data were obtained on the S-II PMN crystal aligned in the (HK0) scattering plane.}
\label{dispersion_100}
\end{figure}

Fig.~\ref{dispersion_100} displays reciprocal space maps, analogous to those in Fig.~\ref{dispersion_110}, that span the (100) Brillouin zone and show the TA$_1$ phonon dispersion along [100] at 600\,K and 300\,K.  The constant inelastic scattering intensity contours in panel $(a)$ reveal a distinct and well-defined TA$_1$ phonon branch at 600\,K.  The same contours in panel $(b)$ show that the TA$_1$ phonon dispersion does not change dramatically and remains well-defined at large $q$ near the X-point zone boundary ($\vec{Q}=(0.5,1,0)$).  This behavior contrasts with that of the TA$_2$ phonon, which becomes overdamped at all $q$ at 300\,K except very close to the zone center. The data in panel $(b)$ also show that the energy of the TA$_1$ mode does not shift significantly between 600\,K and 300\,K at the zone boundary.  There is, however, a strong indication that some extra broadening of the TA$_1$ phonon is present for wavevectors near the zone center at 300\,K.

\begin{figure}[t]
\includegraphics[width=80mm]{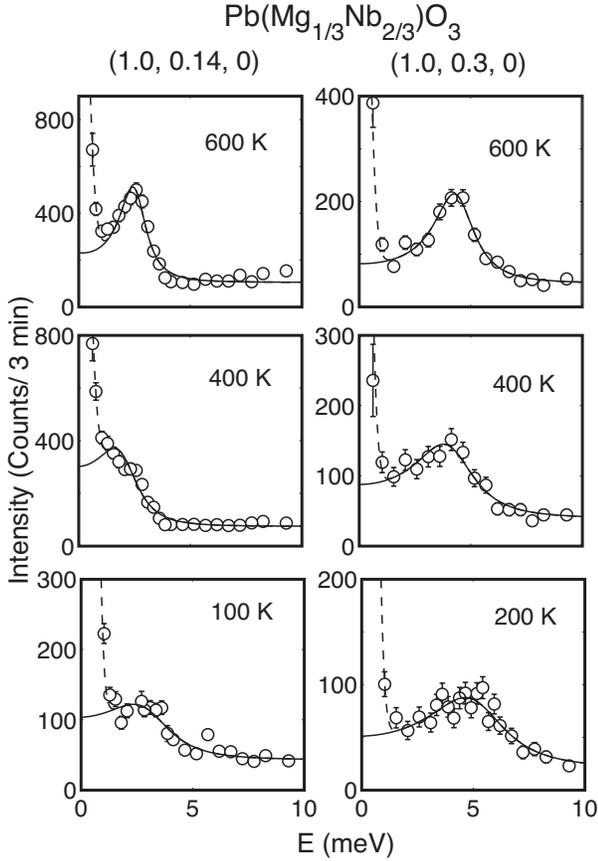}
\caption{Constant-$\vec{Q}$ scans measured at $\vec{Q}=(1.0,0.14,0)$ and $(1.0,0.3,0)$ at 600\,K, 400\,K, 200 \,K,and 100\,K.  The solid lines represent fits to a damped harmonic oscillator as described in the text.  The data were taken with the C5 triple-axis spectrometer located at Chalk River Laboratory using $E_f=14.6$\,meV.  The PMN S-I sample was aligned in the (HK0) scattering plane.}
\label{T1_allT}
\end{figure}

To explore the wavevector and temperature dependence of the damping of the TA$_1$ phonon in more detail, we examine the series of constant-$\vec{Q}$ scans shown in Fig.~\ref{T1_allT}.  At 600\,K, which is above both $T_C$ and $T_d$, well-defined and underdamped TA$_1$ phonons are observed.  At 400\,K, which is near $T_d = 420$\,K, i.\ e.\ the temperature where elastic diffuse scattering first appears and where the TO phonon energy reaches a minimum value, the TA$_1$ phonons at small wavevector ($q=0.14$\,rlu) become overdamped, which is consistent with the behavior suggested by the contours in Fig.~\ref{dispersion_100}.  The TA$_1$ phonons at larger wavevectors ($q=0.3$\,rlu) display some evidence of a broadened linewidth, however they remain underdamped.  This is a key point of this work:  the damping of the TA$_1$ phonon measured at $\vec{Q}=(1,0.3,0)$ is not nearly as dramatic as that observed for the TA$_2$ mode at a comparable wavevector measured along [1$\overline{1}$0] in the (110) Brillouin zone (see the constant-$\vec{Q}$ scans measured at $\vec{Q} = (1.25,0.75,0)$ in Fig.~\ref{T2_allT}).  At temperatures below $T_C$ the TA$_1$ lineshape recovers, but the phonon is still heavily damped at small $q$.

\begin{figure}[t]
\includegraphics[width=90mm]{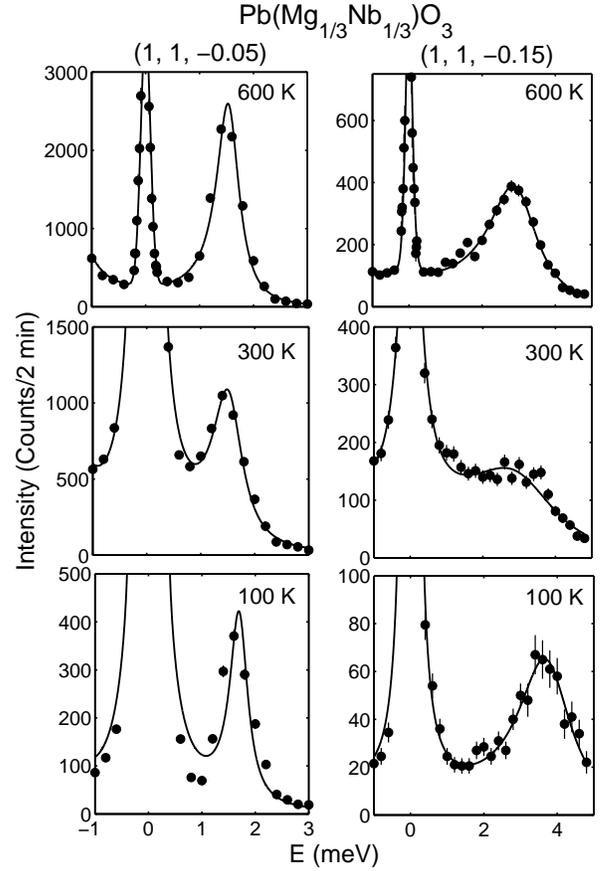}
\caption{Constant-$\vec{Q}$ scans measured at $\vec{Q}=(1.0,1.0,-0.05)$ and $(1.0,1.0,-0.15)$ at 600\,K, 300\,K, and 100\,K.  The solid lines represent fits to a damped harmonic oscillator plus a Lorentzian function centered at $E=0$.  The data were taken with the SPINS cold-neutron, triple-axis spectrometer located at NIST using $E_f=4.5$\,meV.  The PMN S-II sample was aligned in the (HHL) scattering plane.}
\label{T1_spins}
\end{figure}

The behavior of the TA$_1$ phonons measured in the (100) Brillouin zone is intermediate to that of TA$_2$ phonons measured in the (110) Brillouin zone and to that of the TA phonons measured in the (200) and (220) Brillouin zones.  Both TA$_1$ phonons in the (100) Brillouin zone and TA$_2$ phonons in the (110) Brillouin zone become overdamped on cooling to temperatures near $T_d$ (but above $T_C$) for a range of wavevectors between the zone center and zone boundary.  However for the TA$_2$ phonons this damping extends all the way to the zone boundary; for the TA$_1$ phonons it does not.  The data presented in Fig.~\ref{dispersion_100} show that TA$_1$ phonons measured at and close to the zone boundary $\vec{Q} = (1,\pm 0.5,0)$ remain underdamped at 300\,K (i.\ e.\ below $T_d$).  At the other extreme is the behavior of TA phonons in the (200) and (220) Brillouin zones, for which overdamped lineshapes are \textit{never} observed at \textit{any} temperature or wavevector.

The low-$q$ damping of the TA$_1$ phonons observed in the (100) Brillouin zone is consistent with the notion of a coupling to the diffuse scattering centered at $E=0$, and Fig.~\ref{summary_100} demonstrates that a significant diffuse scattering cross section is present near (100) at small wavevectors measured along [010].  While we have examined the TA$_1$ phonons for wavevectors $q \geq 0.14$\,rlu, it is important to investigate TA$_1$ phonons at wavevectors in the long-wavelength ($q \rightarrow 0$) limit.  It was shown in Ref.~\onlinecite{Gehring09:79} that TA$_2$ phonons measured in the (110) Brillouin zone at $q=0.10$\,rlu exhibit a strong damping that persists to the Brillouin zone boundary; this finding is consistent with the data in Fig.~\ref{dispersion_110}.  However TA$_2$ phonons measured at much smaller $q=0.035$\,rlu do not show the dramatic changes in linewidth and energy that are seen at larger $q$.   We observe a similar behavior for the long-wavelength TA$_1$ phonons measured in the (100) Brillouin zone.  Data taken with good energy and $q$-resolution are presented in Fig.~\ref{T1_spins}.  These data were measured in the (HHL) scattering plane near the $\vec{Q}= (1,1,0)$ and display TA$_1$ phonons at $q=0.05$\,rlu and 0.15\,rlu that can be compared to those presented in Fig.~\ref{T1_allT}.  While a strongly damped TA$_1$ lineshape is confirmed for $q=0.15$\,rlu at 300\,K, which recovers upon either heating to 600\,K or cooling to 100\,K, the TA$_1$ lineshape at the smaller wavevector of $q=0.05$\,rlu shows comparatively little change in linewidth and almost no shift in energy.  We therefore conclude that both TA$_1$ and TA$_2$ phonons are insensitive to the formation of polar nanoregions in the long-wavelength limit.  The essential difference is that the TA$_2$ phonons measured in the (110) Brillouin zone are strongly damped all of the way to the Brillouin zone boundary whereas the TA$_1$ phonons are strongly damped only for an intermediate range of momentum transfer $q$.

\section{COMPARISON BETWEEN DIFFERENT BRILLOUIN ZONES}

\begin{figure}[t]
\includegraphics[width=80mm]{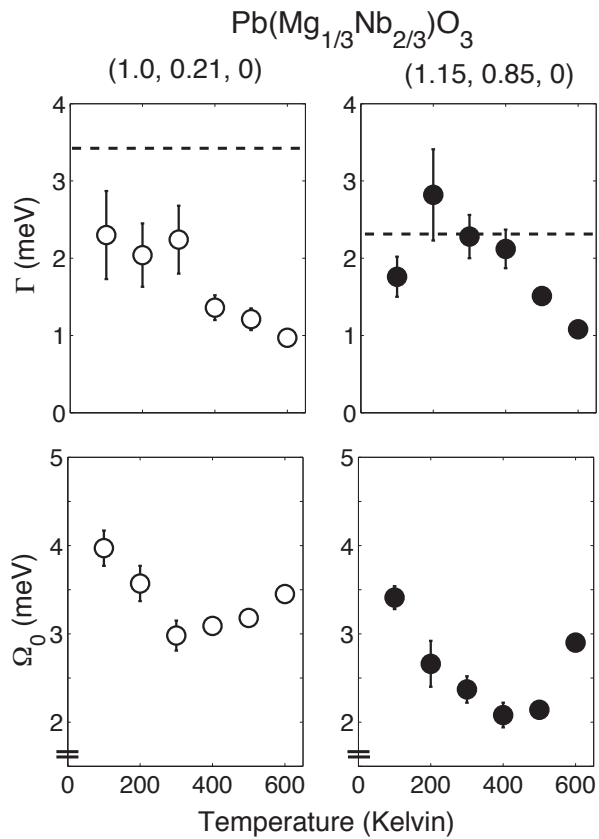}
\caption{The temperature dependence of the linewidth $\Gamma$ (upper panels) and energy $\Omega_0$ (lower panels) of the TA$_1$ and TA$_2$ phonons was determined by fitting a damped harmonic oscillator lineshape to constant-$\vec{Q}$ scans measured at $\vec{Q}=(1.0,0.21,0)$ and $(1.15,0.85,0)$, respectively.  The horizontal dashed lines in the two upper panels represent the average TA phonon energy and thus indicate when the phonon becomes overdamped ($\Gamma \geq \Omega_0$).  The error bars are derived from the least-squares fit to the data.}
\label{proc_strong_diffuse}
\end{figure}

\begin{figure}[t]
\includegraphics[width=80mm]{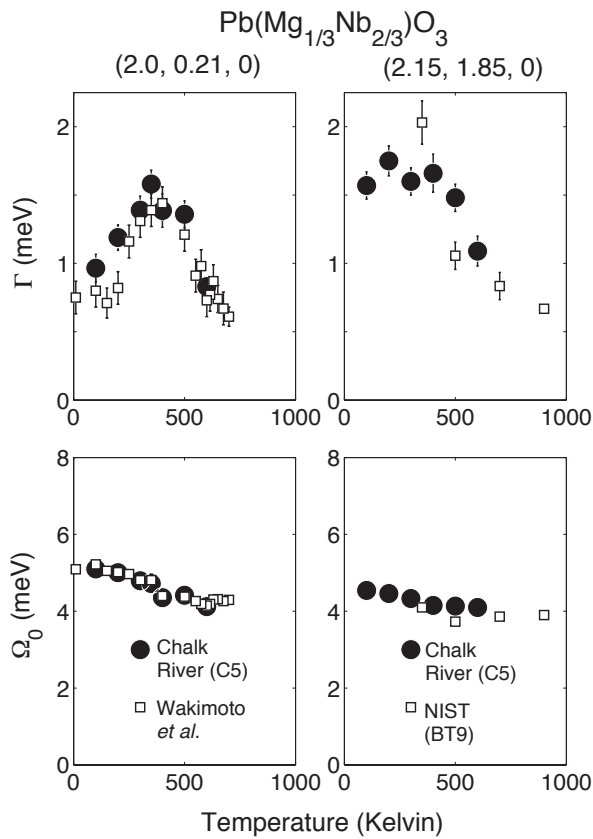}
\caption{The temperature dependences of the linewidth $\Gamma$ (upper panels) and energy $\Omega_0$ (lower panels) of the TA$_1$ and TA$_2$ phonons measured in the (200) and (220) Brillouin zones, respectively.  The values are obtained from constant-$\vec{Q}$ scans measured at the exactly same reduced wavevector $\vec{q}$ as those shown in Fig.~\ref{proc_strong_diffuse}.  The error bars are derived from the least-squares fit to the data; for the energy positions the error bars are equal to the size of the data points.}
\label{proc_weak_diffuse}
\end{figure}

The results from the previous section can be summarized in the following two figures in which the temperature dependences of the linewidth ($\Gamma$) and energy position ($\Omega_0$) are plotted for TA$_1$ and TA$_2$ phonons measured at the same $|\vec{q}| = \sqrt{0.15^2 + 0.15^2} \sim 0.21$\,rlu over a large temperature range.  Figure~\ref{proc_strong_diffuse} presents the lineshape analysis of the TA phonons measured in the (100) and (110) Brillouin zones, where the diffuse scattering cross section is strong, while Fig.~\ref{proc_weak_diffuse} presents that for TA phonons measured in the (200) and (220) Brillouin zones, where the diffuse scattering cross section is weak.

We start with the (100) and (110) Brillouin zones.  The upper panels of Fig.~\ref{proc_strong_diffuse} show how the linewidth of the TA phonons varies from 100\,K to 600\,K, and the dashed lines represent the average TA phonon energy position over the same temperature range.  These lines therefore represent the threshold above which the TA phonon lineshape changes from being underdamped ($\Gamma \textless \hbar \Omega_{0}$) to overdamped ($\Gamma \geq \hbar \Omega_{0}$).  From this one can immediately see that the TA$_1$ phonon at this wavevector, although heavily damped, remains underdamped at all temperatures.  By contrast, the TA$_2$ phonon at the same $|\vec{q}|$ becomes overdamped between 200\,K and 400\,K.  We emphasize that this does \textit{not} imply that the TA$_1$ phonon is never overdamped; on the contrary, the constant-$\vec{Q}$ scan measured at lower $q$ $(1,0.14,0)$ shown in Fig.~\ref{T1_allT} reveals a clearly overdamped TA$_1$ lineshape at 400\,K.  The point of this analysis is to show that for a given temperature the overdamping of the TA$_2$ phonon persists to larger $q$ than does that for the TA$_1$ mode.  Hence the damping mechanism affecting the TA modes in PMN is anisotropic in reciprocal space, and this anisotropy correlates with that of the diffuse scattering cross section.

The lower two panels of Fig.~\ref{proc_strong_diffuse} show that the TA phonon energy positions shift significantly with temperature.  Of particular note is that the TA mode energies are both minimal at a temperature that coincides well with T$_{d} \sim$ 420\,K (where elastic diffuse scattering first appears, see Fig.~\ref{diffuse_cte}, and the TO phonon energy is lowest).  However these data also reveal a key difference in that the TA$_2$ phonon energy position is significantly lower than that for the TA$_1$ phonon over the entire temperature range.

Having demonstrated that the TA phonons in PMN are strongly coupled to the diffuse scattering cross section, we revisit the (200) and (220) Brillouin zones where the elastic diffuse scattering cross section is weak, but non-zero. To this end we show in Fig.~\ref{proc_weak_diffuse} the temperature dependence of the linewidth ($\Gamma$) and energy positions ($\Omega_0$) of the \textit{same} phonons studied in Fig.~\ref{proc_strong_diffuse}, but measured in the (200) and (220) Brillouin zones.  Fig.~\ref{proc_weak_diffuse} also combines previously published and unpublished data measured on other PMN single crystals using the BT9 thermal-neutron, triple-axis spectrometer, located at NIST, with data taken on the larger S-I and S-II PMN crystals studied here.  The agreement between these data sets prove that there is no sample dependence to the results discussed in this paper.  They also demonstrate that both the TA$_1$ and TA$_2$ phonons measured in these zones are strongly damped.  However, unlike the behavior observed in the (100) and (110) Brillouin zones, these TA phonons are never overdamped.  Moreover, the TA phonon energy positions display conventional behavior in that they slowly increase with decreasing temperature.  We also observe a key difference wherein the TA$_1$ phonon linewidth narrows markedly below $T_d$ while the TA$_2$ phonon remains strongly damped to lowest temperatures.  It is interesting to note that the largest TA phonon linewidths are roughly comparable to those measured in the (100) and (110) Brillouin zones; however the TA phonons in the (200) and (220) Brillouin zones never become overdamped because the energy positions are not affected to the same degree as they are in the (100) and (110) Brillouin zones.

The lower-$q$ TA phonons display strong damping below $T_d$, and it is only below $T_C$ that the TA$_1$ phonons recover; by contrast the damping of the TA$_2$ phonons persists to low temperature.  This result is consistent with the highly anisotropic correlations proposed in several models involving domain walls or short-range, polar correlations existing along [1$\overline{1}$0] as TA phonons propagating along this direction will be most disrupted by the disorder.  The Brillouin zone dependence of the TA phonon energy position that results in a strongly overdamped lineshape in the (110), but not in the (220) Brillouin zone comes about from the coupling between the acoustic mode and the diffuse scattering cross section.

The diffuse scattering in PMN has recently been shown to have an inelastic component that was directly observed in neutron spin-echo measurements~\cite{Stock10:81} and that was proposed on the basis of cold-neutron scattering experiments in Ref.~\onlinecite{Gvas04:49}.  In these experiments it was found that by fitting the spectrum the inelastic linewidth of the diffuse scattering cross section varied as $q^2$.  The energy scale was found to be comparable to the TA phonon energy, thus providing a strong channel for coupling and renormalization of the phonon energy.   The diffuse scattering cross section along the direction of propagation of TA$_1$ phonons in the (100) Brillouin zone is not as large as that for TA$_2$ phonons in the (110) Brillouin zone.  Since the energy scale and intensity of the diffuse scattering along $\langle 100 \rangle$ are presumably much weaker, it seems logical that it would only provide a coupling over a small range of wavevectors where the energy scale matches that of the corresponding TA$_1$ phonon.  The apparent lack of coupling at the smallest wavevectors, as illustrated in Fig.~\ref{T1_spins}, would then result from the energy scale of the diffuse scattering being too small to strongly couple to the TA phonon.  This physical model of coupling is consistent with the general ideas proposed for mode coupling in KTaO$_3$ (Ref.~\onlinecite{Axe70:1}) where it was suggested that the coupling constant scales linearly with the reduced wavevector $q$ measured relative to the Brillouin zone center.

It is important to compare our neutron scattering measurements of the low-energy acoustic phonons in PMN with those using other techniques including Brillouin scattering and ultrasound.  Ref.~\onlinecite{Lushnikov08:77} presents a summary of both Brillouin (measured by the authors) and ultrasound measurements (taken from Ref.~\onlinecite{Smo85:27}) covering a broad range of frequencies spanning from the MHz to GHz range.   Their data demonstrate a softening of the elastic constants at temperatures below 300\,K, and the magnitude of the softening is frequency dependent.  The linewidth of the Brillouin peaks has also been reported in Ref.~\onlinecite{Tu95:78}, and a broadening and recovery have been reported for the longitudinal acoustic (LA) [001] mode, but no broadening was observed for the transverse mode.  These results are generally inconsistent with the results reported here.  We observe significant broadening for the TA$_1$ and TA$_2$ phonons with the former displaying a recovery at low temperatures.  We also observe that the anomalies in the energy position become less pronounced at smaller wavevectors and lower energies.  We remark that our data reflects the lattice dynamics of PMN on the THz energy scale, which is nearly three orders of magnitude larger than that reported by ultrasound or Brillouin scattering experiments.  We therefore conclude that the effects reported at much lower frequencies are distinct from those presented here.  We also note that infrared and dielectric experiments have reported significant dynamics down to much lower frequencies than are accessible with neutron scattering techniques.

\section{ELASTIC CONSTANTS OF PMN}

Based on the slopes of the acoustic phonon dispersions near the zone center, we can calculate the elastic constants using the equations of motions outlined in Ref.~\onlinecite{Dove:book}.  For a cubic crystal, the elastic constant matrix is determined by three values - C$_{11}$, C$_{44}$, and C$_{12}$.  C$_{44}$ is determined uniquely from the slope of the TA$_1$ mode and C$_{11}$ is fixed by longitudinal modes polarized along [100].  The elastic constant C$_{12}$ is not uniquely determined, however the slope of the TA$_2$ mode is determined by the difference (C$_{11}$-C$_{12}$), and the slope of the longitudinal mode polarized along [110] is set by the sum (C$_{11}$+C$_{12}$+2C$_{44}$).~\cite{Noda89:40}  Therefore, C$_{12}$ can be determined from our measurements once C$_{11}$ and C$_{11}$-C$_{12}$ are known, and it can be checked for consistency between the two branches.

\begin{table}[h!]
\begin{tabular}{c|c|c|c|}
 Material (Technique)	& 		C$_{11}$  	& 		C$_{12}$ 		& 	C$_{44}$\\
 \hline
\hline
PMN(neutron) 		    & 		$1.39(7)$		& 		$0.43(5)$ 		& 	$0.53(3)$  \\
PMN(Brillouin-Tu \textit{et al.}) 			& 		1.49			& 		- 			& 	0.68 \\
PMN(Brillouin-Ahart \textit{et al.}) 			& 		1.56			& 		0.76 			& 	0.685 \\
\hline
\hline
PT (neutron-Tomeno \textit{et al.}) 			& 		2.32			& 		1.06 			& 	0.72 \\
PT (Brillouin-Li \textit{et al.}) 			& 		2.35			& 		0.97 			& 	0.65 \\
\hline
 \hline
\end{tabular}
\caption{A summary of the elastic constants (in units of $10^{11} N/m^2$) measured at 300\,K for a series of PMN and PT compounds.  The Brillouin and neutron scattering results on PT were taken from Ref.~\onlinecite{Tu95:78,Ahart07:75,Tomeno06:73,Li93:41}. The elastic constants were calculated assuming a density $\rho$=8.19 g/cm$^{3}$.}
\label{table:sample}
\end{table}

We have determined the elastic constants by measuring the slopes of the transverse and longitudinal acoustic phonons near the (200) and (220) Brillouin zone centers.  We have chosen these zones because the diffuse scattering cross section is very weak and the temperature dependence of the phonon energy positions exhibits little effects from the coupling to the relaxational dynamics that are prevalent in the (100) and (110) Brillouin zones.  To extract an accurate value for the limiting slope for the phonon dispersion curves we have restricted ourselves to wavevectors less than 15\% of the Brillouin zone in order to assure linearity in the dispersion.

The values of the elastic constants for PMN at 300\,K are listed in Table~\ref{table:sample} and are compared to values obtained from Brillouin scattering measurements.  The neutron and Brillouin scattering measurements agree well and the discrepancy can be resolved through the fact that neutrons probe the dynamics on the THz timescale where Brillouin techniques study the dynamics in a significantly lower frequency range. The published values for PbTiO$_3$ are also listed for comparison.  While C$_{44}$ for PMN and PbTiO$_3$ are very similar, both C$_{11}$ and C$_{12}$ are reduced in PMN.  The velocity of the TA$_2$ mode (set by C$_{11}$-C$_{12}$) is also reduced with C$_{11}$-C$_{12}$=0.88 for PMN and 1.32 for PbTiO$_3$.   It is interesting to note that near the morphotropic phase boundary (MPB), which occurs when $\sim 33$\%PT  is substituted for PMN, the difference C$_{11}$-C$_{12}$ reduces further and nearly approaches zero, but then increases for Ti-rich compositions beyond the MPB boundary.~\cite{Cao04:96}  This suggests the presence of an instability of relaxors to TA$_2$ phonons.~\cite{Cowley13:13}  This softening of the acoustic branch maybe a contributing factor to the increased coupling observed between the relaxational and harmonic modes in the (100) and (110) Brillouin zones.  Another difference is seen in the values of the elastic anisotropy, defined as 2C$_{44}$/(C$_{11}$-C$_{12}$), which for PMN is 1.43 and for PbTiO$_3$ is 0.75.  Again, this is indicative of a softening of the TA$_2$ branch and possibly indicative of an instability to a homogeneous deformation along $\langle 110 \rangle$.

Fig.~\ref{proc_weak_diffuse} summarizes the energy positions of the TA$_1$ and TA$_2$ modes as a function of temperature in the (200) and (220) Brillouin zones.  These data show that a small increase occurs with decreasing temperature, but there is no concurrent softening of the elastic constants.  The data are suggestive that, on the THz timescale probed by neutrons, there are only relatively modest changes in the elastic constants.  This finding stands in contrast to that in Ref.~\onlinecite{Lushnikov08:77}, which presents Brillouin data measured  in the MHz range that shows a large softening of the elastic constants (including C$_{44}$).  We emphasize that our experiments are limited by the energy resolution of the spectrometer to the THz energy range, and we are not able to access the MHz region presented in Ref.~\onlinecite{Lushnikov08:77}.  As discussed in Ref.~\onlinecite{Cowley67:90}, the difference in elastic constants at these extremes in frequency maybe due to anharmonic effects.  Dielectric and infrared measurements have shown significant dynamics at low energies, which might be contributing to the apparent discrepancy between the MHz and THz range.~\cite{Bovtun09:79}

\section{COMPARISON BETWEEN THE RELAXOR PMN AND THE FERROELECTRIC PMN-60\%PT}

 \begin{figure}[t]
\includegraphics[width=75mm]{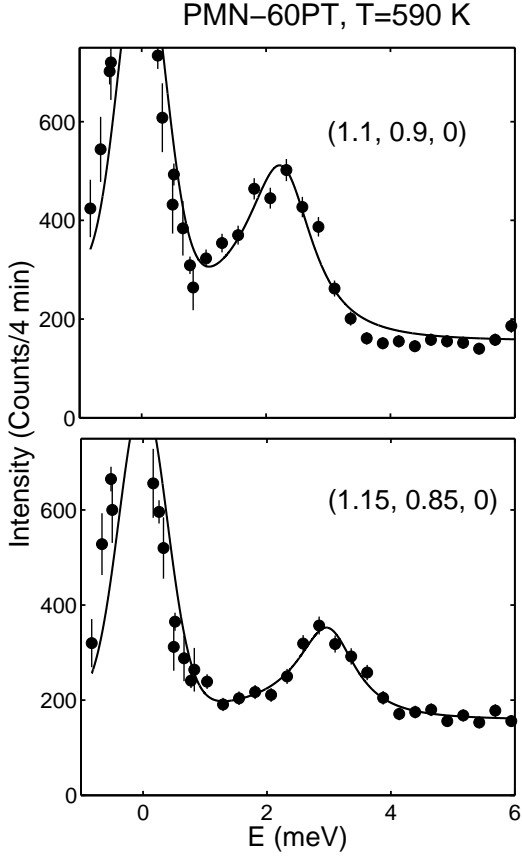}
\caption{The TA$_2$ phonon in PMN-60\%PT measured in the (110) Brillouin zone.  The data were taken using the N5 thermal-neutron, triple-axis spectrometer with the sample aligned in the (HK0) scattering plane.}
\label{PMN_60PT_figure}
\end{figure}

As demonstrated in Ref.~\onlinecite{Mat06:74}, the diffuse scattering cross section can be tuned through PbTiO$_3$ (PT) doping, and a suppression of the diffuse scattering cross section is observed for large dopings that lie beyond the MPB.  Such Ti-rich compositions also show no relaxor properties; instead they exhibit well-defined ferroelectric phases at low temperatures similar to that reported in PbTiO$_3$.  As further evidence that the TA phonon damping reported here in PMN is related to the polar nanoregions that are associated with the diffuse scattering cross section, we present several comparative constant-$\vec{Q}$ scans of the TA$_2$ phonons in PMN-60\%PT measured near the (110) Brillouin zone in Fig.~\ref{PMN_60PT_figure}.  PMN-60\%PT displays no measurable diffuse scattering and displays a well-defined, first-order phase transition from a cubic to a tetragonal unit cell.  Concomitant with this transition the zone center TO mode softens and then hardens in a manner similar to that measured in pure PbTiO$_3$.~\cite{Shirane70:2,Hlinka06:73}   The phonons in Fig.~\ref{PMN_60PT_figure} are also well-defined and underdamped.  The temperature of 590\,K at which these scans were made was chosen because it is just above the structural transition of $\sim 550$\,K and serves to demonstrate that the suppression of the diffuse scattering cross section removes the strong damping and energy renormalization reported earlier for PMN in Fig.~\ref{dispersion_110}.  These results confirm the fact that the large damping and softening of the TA$_2$ mode in pure PMN is the result of short-range, polar correlations or polar nanoregions.

\section{DISCUSSION AND CONCLUSIONS}

\begin{figure}[t]
\includegraphics[width=85mm]{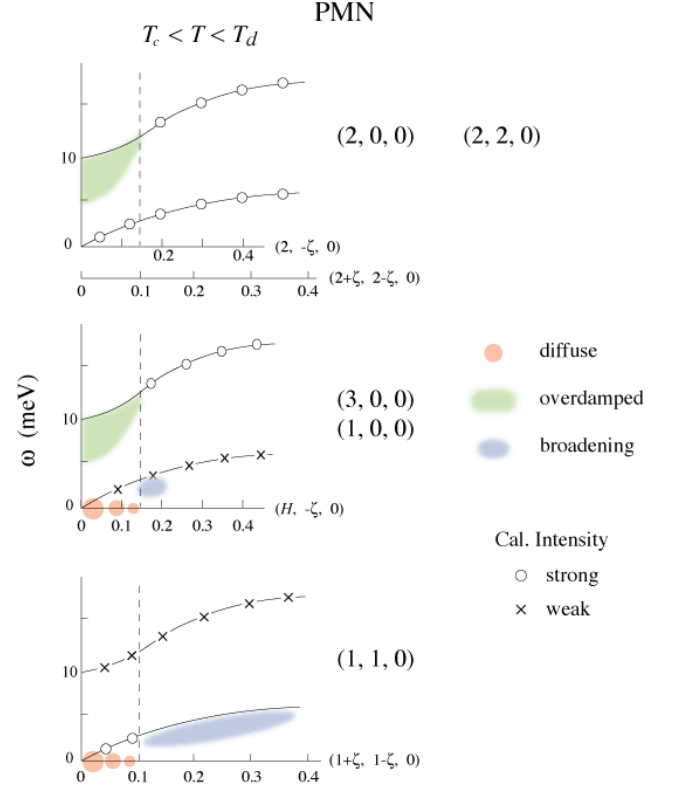}
\caption{A schematic illustration of the phonon dynamics for the TA$_1$ and TA$_2$ phonons measured in Brillouin zones where the diffuse scattering is strong ((100), (110), and (300)) and weak (200).  The TO mode in the (110) zone was too weak to provide any information on the dampening.  While the TO mode in the (100) zone was also prohibitively weak, the diagram above assumes the lineshape should scale as the structure factor. Further discussion of the curves is provided in the text.}
\label{cartoon}
\end{figure}

We have reported the results of a detailed study of the acoustic phonons in a number of Brillouin zones in PMN.  We have observed that the TA phonons measured in the (200) and (220) Brillouin zones are well-defined and exhibit a conventional temperature dependence expected from harmonic theory in the absence of strong TA-TO mode coupling.   While the TO phonon cross section in these zones is strong and readily observable, the diffuse scattering intensity is weak.  The situation in the (100) and (110) Brillouin zones is completely different.  In these zones the diffuse intensity is strong and the optic phonon structure factor is weak.   While the TA phonons in these zones (particularly in the (110) Brillouin zone) take on a heavily damped lineshape because of a strongly renormalized energy scale, the energy linewidth is similar to that observed in the (200) and (220) Brillouin zones.  Because the lattice dynamics in different Brillouin zones is invariant, we attribute the damped lineshape in these zones to a strong coupling between the diffuse scattering intensity and the TA mode.

For temperatures $T_{c}\textless T \leq T_{d}$, the TA$_2$ phonons studied in the (110) Brillouin zone become overdamped over nearly the entire zone.  However, low-$q$ phonons representing long-wavelength excitations remain underdamped as expected on the basis of a coupling strength that scales with $q$.~\cite{Axe70:1}  The TA$_1$ phonons measured in the (100) Brillouin zone display significant damping at small wavevectors.  These phonons recover below T$_{c}$.  At larger wavevectors, however, while some damping is seen, the TA$_1$ phonons remain underdamped at all temperatures.  The difference in behavior between TA$_{1}$ and TA$_{2}$ phonons is consistent with the idea of coupling between the diffuse scattering cross section and TA phonons because the former is a maximum when measured along [1$\overline{1}$0].

The behavior of the low-energy TA and TO phonons is qualitatively summarized in Fig.~\ref{cartoon}.  This figure illustrates the low-energy TA and TO modes and denotes a broadening of the mode, indicative of damping, through the shading.  Based on our experimental measurements, we observe a direct correlation between the damping of the TA mode and the the diffuse scattering cross section; we therefore conclude that there is a strong coupling between the two.  We have also reported strong damping of the TA phonon in the (200) and (220) Brillouin zones, representing the anisotropy of the polar correlations.

Indeed the presence of a strong relaxational component to the diffuse scattering may help to explain recent hyper-Raman data,~\cite{Zein10:105} which were used to support the case for a second, low-energy TO mode.  We note that this study never observed a well-defined, underdamped mode, but rather a strongly-damped feature that could be interpreted in terms of relaxational dynamics.  Thus many of the anomalous features resulting from extra intensity at low energies in the (300) Brillouin zone could be interpreted in terms of the strong diffuse scattering coupling with the harmonic modes as observed in our data, which were measured in the absence of strong optic modes.

\subsection{The Origin of the Waterfall Effect is not due to TA-TO mode coupling}

We do not agree with the interpretation of the waterfall effect, or the temperature dependence of the TA and TO modes, in terms of coupling between TA and TO modes.  Such suggestions have been motivated by a comparison with BaTiO$_3$.~\cite{Hlinka03:91}  However, the lattice dynamics of BaTiO$_3$ and PMN are completely different.  The coupling in BaTiO$_3$ along [110] results from the fact that the TA and soft TO mode branches cross at small wavevectors; thus, because these modes also have the same symmetry, they become strongly coupled and spectral weight is transferred between the modes, which then produces large energy shifts.~\cite{Shirane70:2}  Such a situation does not occur in PMN because the lattice dynamics are relatively isotropic in comparison to BaTiO$_3$ and no mode crossing occurs.  We have shown that no significant changes occur in either the TA phonon energy position or the spectral weight, both of which should happen in a strong coupling picture.  While some small amount of coupling must exist owing to the fact that the low-energy TA and TO phonons have the same symmetry, we do not believe that this weak coupling dominates the physics of PMN as previously suggested based on the comparison with BaTiO$_3$.

We do see anomalies in the TA phonons measured in the (100) and (110) Brillouin zones.  But we have proven that these anomalies cannot be the result of coupling to the TO mode because they are absent in the (200) and (220) Brillouin zones, precisely where the TO phonon structure factor is strong.  By contrast, the TO phonon structure factor is weak in the (100) and (110) Brillouin zones and this phonon is not easily seen in a constant-$\vec{Q}$ scan.  This makes the TA-TO mode coupling idea highly improbable because the coupling comes from terms proportional to $\lambda F_{TA} F_{TO}$, where $\lambda$ is the coupling constant and $F$ is the phonon structure factor.~\cite{Harada71:4}  A Brillouin zone in which the TO mode cross section is weak will therefore display weak coupling effects between TA and TO phonon branches.  Interpreting the phonon anomalies we have observed in (100) and (110) Brillouin zones in terms of TA-TO mode coupling is therefore not self consistent.

Several studies, including Ref.~\onlinecite{Hlinka03:91} and Ref.~\onlinecite{Waki02:66}, have proposed TA-TO mode coupling in order to explain anomalies observed at small wavevectors in the (300) Brillouin zone.  But this zone has a very strong diffuse scattering cross section (see Ref.~\onlinecite{Hirota02:65}) and a strong TO mode, which softens to low energies at small wavevectors.  It is possible that a coupling may exist between the diffuse scattering and the TO mode, and this might provide an alternate explanation of some of the anomalous features in this zone such as the observation of a different $q$-dependent TO phonon damping compared to other Brillouin zones (Ref.~\onlinecite{Hlinka03:91}) and an apparent enhancement of the TA mode intensity (Ref.~\onlinecite{Waki02:66}).  It might also provide an explanation for the presence of a second ``quasi-optic" mode that was also observed in the (300) Brillouin zone.~\cite{Vak02:66}

We believe a much more likely cause of the broadening of the soft TO mode (i.\ e.\ the waterfall effect) may be the presence of strong disorder induced by random, dipolar fields.  This would explain the broadening observed in both pure PMN and PMN-60\%PT, which is largely absent in the parent compound PbTiO$_3$.  It is interesting to note that disorder has been suggested as the origin of the broadening of the soft TO mode in BaTiO$_3$.~\cite{Dove:book}

\subsection{A second low-energy optic mode as the cause for the anisotropic dynamics?}

The broadening of the low-energy transverse optic mode has been explained in terms of a second mode based on hyper-raman (Ref. \onlinecite{Zein10:105}) and some neutron scattering results (Refs. \onlinecite{Vak02:66,Vak10:52}).  We provide an alternative explanation in this paper for the low-energy spectral weight and we do not believe that our results, particularly in the $\vec{Q}$=(1,1,0) and (1,0,0) zones, are consistent with this picture.  The broadening that we observe at low energies in these Brillouin zones becomes less pronounced and to vanish in the limit $q \rightarrow$ 0.  This is consistent with coupling to a relaxational component and inconsistent with a zone center feature such as a second optic mode.  Furthermore, we note that the broadening coincides with T$\sim$ 400 K where a static component to the diffuse scattering is onset at lower temperatures (Ref. \onlinecite{Stock10:81}), therefore linking the anomalous broadening we observe to the relaxational dynamics characterizing the diffuse scattering. We note that we have not pursued an understanding of the $\vec{Q}$=(3,0,0) Brillouin zone (as discussed in Refs. \onlinecite{Vak02:66,Vak10:52}) in this report as the combination of strong optic, acoustic, and diffuse scattering are present and will undoubtably complicate the interpretation of the dynamics.

Hyper-raman experiments (Ref. \onlinecite{Zein10:105}), however, illustrate the presence of significant spectral weight at small energy transfers consistent with a second low-energy optic mode.  The temperature dependence was believed to point to a softening at 400 K, consistent with the neutron results described in Refs. \onlinecite{Vak02:66,Vak10:52}.  The interpretation in terms of two modes is further corroborated by the splitting of higher energy optic modes probed using Raman spectroscopy.~\cite{Zein11:109,Zein08:89,Hehlen07:75}  The low-energy measured peaks, however, are significantly broader than the resolution and are over damped with the linewidth being larger or at least comparable with the energy position.  We therefore believe that the data is consistent the results presented here in the $\vec{Q}$=(2,0,0) and (2,2,0) zones where the zone center scattering is interpreted in terms of a single over damped mode.  The hyper raman results maybe also consistent with the relaxational component to the diffuse scattering observed with spin-echo techniques.  The comparison made by Raman techniques to higher energy split modes is compelling and neutrons have not yet been successful in observing a splitting of these higher energy modes. 

\subsection{Anisotropic polar nanoregions}

Several models of the diffuse scattering have challenged the idea that polar nanoregions are highly anisotropic in real space.~\cite{Bosak11:xx}  The results presented here challenges these ideas and instead point to a damping mechanism that is highly anisotropic given the large differences observed between the TA$_1$ and TA$_2$ phonons in the (HK0) scattering plane.  The large degree of anisotropy in the damping implies some sort of real space structure that is also highly anisotropic.  Phonons propagating along directions that exhbiti weak polar correlations are likely to be more heavily damped than phonons propagating along directions where long range correlations exist.  Our results further suggest that short-range, polar order is responsible for the strong damping of the TA phonons and that the size of this damping is tied to the correlation length associated with the diffuse scattering.  We note that the TA phonon damping reported here only reflects anisotropic correlations.  Thus we are not able to favor the ``pancake" model over the [1$\overline{1}$0] domain model proposed by different groups.  The results here simply point to anisotropic polar correlations.

\subsection{Dynamics, two temperature scales, and the high-temperature scale $T_d$}

Considerable debate and discussion have been devoted to the nature and existence of the well-known Burns temperature, which for PMN is purported to occur around 620\,K.  Our studies here suggest that this high-temperature scale is a purely dynamic phenomena and thus depends on the energy resolution of the experimental probe used to measure it.  This is consistent with results from neutron dynamic pair density function analysis, which illustrate that the off-centering of the lead in PMN is dynamic at temperatures above $\sim 350-400$\,K.~\cite{Dmowski08:100}  Static, polar regions begin to appear at 420\,K where we observe a minimum in the soft TO frequency and the appearance of static diffuse scattering on GHz timescales measured with neutron inelastic scattering.  This lower temperature is also consistent with Raman studies in which a splitting of some modes was observed, although this was interpreted as evidence for a third temperature scale.~\cite{Tolouse08:369}   A third temperature scale has also been postulated based lattice constant measurements;~\cite{Dkhil09:80} however, as demonstrated here, such evidence is highly suspect given that the thermal expansion is sample dependent, and possibly results from the effect of a significant skin effect.

We believe that all of the results on PMN can be broadly interpreted in terms of only two temperature scales.  A high temperature scale ($T_d \sim 420$\,K), which is characterized by the appearance of static polar correlations and where a minimum in the soft mode energy exists, and a lower temperature scale ($T_C \sim 210$\,K), where domains form and a long-range ferroelectric phase can be induced through the application of a sufficiently strong electric field.  We believe the Burns temperature is dynamical and should not be interpreted in terms of a well-defined phase transition.  This idea was discussed in Ref.~\onlinecite{Gehring09:79} where the Burns temperature was redefined to be $T_d = 420 \pm 20$\,K for PMN, instead of the much higher value of 620\,K that had been derived from index of refraction measurements.

The two temperature scales discussed here are naturally reflected in the dynamics of the TA phonons presented above.  The high temperature scale ($T_d$) is where the broadening of both the TA$_1$ and TA$_2$ phonons is a maximum and where the energy renormalization due to the coupling to the relaxational diffuse scattering is most apparent.  The lower temperature scale ($T_C$) marks the recovery of the TA$_1$ phonon in terms of a decrease in the damping in energy.  This lower temperature scale is not strongly apparent in the dynamics of the TA$_2$ phonon, however, we propose that this is due to the real space structure of the low-temperature ferroelectric domains.  We emphasize that our results only demonstrate that the ferroelectric correlations in real space are highly anisotropic; this is consistent with the pancake model (Ref.~\onlinecite{Xu03:70,Welberry05:38,Welberry06:74}) and all other models that invoke anisotropic correlations (for example Ref.~\onlinecite{Pasciak07:76,Ganesh10:81}).  We are not able to distinguish between these various models on the basis of our experimental data.

\subsection{Summary}

We have demonstrated an anisotropic damping and energy renormalization of the TA phonons in PMN.  TA phonons in all zones display strong damping below $T_d$, where static diffuse scattering is onset, indicative of short-range, polar order.  Only below $T_C$, where the diffuse scattering becomes completely static, do TA$_1$ phonons propagating along [100] recover a normal lineshape.  However, the TA$_2$ phonons, which travel along [1$\overline{1}$0], do not, therefore revealing the underlying anisotropic structure of the polar correlations.  TA phonons measured in zones that have a strong diffuse cross section exhibit a strong renormalization in energy, which is indicative of a coupling to the relaxational component of the diffuse scattering.  This coupling might provide an explanation for the extra neutron scattering intensity measured in the (300) Brillouin zone that was used to support the ideas of TA-TO mode coupling and the presence of a second, low-energy, TO mode.  We have also provided measurements of the three elastic constants in the THz energy range.  These are all in reasonable agreement with published data using Brillouin scattering.  When compared to pure PbTiO$_3$, our data suggest an instability exists in PMN that is related to the TA$_2$ acoustic branch.

\section{ACKNOWLEDGMENTS}

We acknowledge financial support from the Natural Sciences and Engineering Research Council of Canada, the National Research Council of Canada, and from the U.S. DOE under contract No. DE-AC02-98CH10886, and the Office of Naval Research under Grants No. N00014-02-1-0340, N00014-02-1-0126, and MURI N00014-01-1-0761.


\end{document}